%% file: main.tex
\pdfoutput=1
\documentclass[10pt,conference]{IEEEtran}
\usepackage[utf8]{inputenc}
\usepackage[T1]{fontenc}
\usepackage{graphicx}
\usepackage{booktabs}
\usepackage{amsmath}
\usepackage{xspace}\xspaceaddexceptions{\%}
\usepackage{url}
\usepackage{xcolor}
\usepackage{tikz}
\usetikzlibrary{positioning}
\usepackage[hidelinks]{hyperref}
\hypersetup{pdftitle={How Python Audio Libraries Change Across Releases: Module Structure and Documentation}, pdfauthor={Mateusz Modrzejewski, Michal Rokita, Lukasz Neumann}}
\graphicspath{{figures/}}

\input{tables/numbers.tex}

\input{tables/numbers_control.tex}

\newcommand{\answer}[2]{\par\smallskip\noindent\fbox{\parbox{\dimexpr\columnwidth-2\fboxsep-2\fboxrule}{\textbf{Answer to RQ#1.} #2}}\par\smallskip}

\begin{document}
\bstctlcite{IEEEexample:BSTcontrol}

\title{How Python Audio Libraries Change Across Releases: Module Structure and Documentation}

\author{\IEEEauthorblockN{Mateusz Modrzejewski, Micha\l{} Rokita, and \L{}ukasz Neumann}
\IEEEauthorblockA{Institute of Computer Science, Warsaw University of Technology, Warsaw, Poland\\
\{mateusz.modrzejewski, michal.rokita, lukasz.neumann\}@pw.edu.pl}}

\maketitle

\input{sections/abstract}
\input{sections/introduction}
\input{sections/design}
\input{sections/results}
\input{sections/discussion}
\input{sections/background}
\input{sections/threats}
\input{sections/conclusion}

\bibliographystyle{IEEEtran}
\bibliography{references}

\end{document}

%% file: tables/numbers.tex
\newcommand{\AddedBN}{7\xspace}
\newcommand{\AddedBNonActive}{4\xspace}

\newcommand{\AddedOtherN}{32\xspace}
\newcommand{\AddedOtherNonActive}{28\xspace}
\newcommand{\AgreeAllThreePct}{70.4\xspace}

\newcommand{\AllORCompanyDormant}{2.7\xspace}

\newcommand{\AuditAdjacentNamedPct}{8.0\xspace}

\newcommand{\AuditDedupNamedPct}{9.0\xspace}

\newcommand{\AuditDedupRestoredUnits}{1\xspace}

\newcommand{\AuditImportableNamedPct}{10.9\xspace}

\newcommand{\AuditImportableUnitsPct}{69.9\xspace}
\newcommand{\AuditLooNamedPctMax}{9.9\xspace}

\newcommand{\AuditLooNamedPctMin}{7.8\xspace}

\newcommand{\AuditNonImportableNamedPct}{4.4\xspace}

\newcommand{\AuditUnitsTopRepoShare}{31.1\xspace}

\newcommand{\AuditUnnamedParentMentioned}{302\xspace}
\newcommand{\AuditUnnamedRemovalWordComponent}{270\xspace}
\newcommand{\AuditUnnamedUnits}{1,103\xspace}

\newcommand{\CliffGrowth}{-0.44\xspace}

\newcommand{\CliffList}{-0.23\xspace}

\newcommand{\CliffSize}{-0.48\xspace}

\newcommand{\CliffUnchanged}{0.05\xspace}

\newcommand{\CtxBodyBreakKappaHi}{0.85\xspace}
\newcommand{\CtxBodyBreakKappaLo}{0.84\xspace}

\newcommand{\CtxBodyChangedHi}{11.7\xspace}
\newcommand{\CtxBodyChangedLo}{6.2\xspace}

\newcommand{\CtxBodyEntryKappaHi}{0.75\xspace}
\newcommand{\CtxBodyEntryKappaLo}{0.63\xspace}

\newcommand{\CtxEntries}{435\xspace}

\newcommand{\CtxFullReproHi}{99.5\xspace}
\newcommand{\CtxFullReproLo}{97.9\xspace}
\newcommand{\CtxHeadingBreakKappaHi}{0.47\xspace}
\newcommand{\CtxHeadingBreakKappaLo}{0.39\xspace}

\newcommand{\CtxHeadingChangedHi}{80.9\xspace}
\newcommand{\CtxHeadingChangedLo}{54.0\xspace}

\newcommand{\CtxHeadingEntryKappaHi}{0.23\xspace}
\newcommand{\CtxHeadingEntryKappaLo}{0.22\xspace}

\newcommand{\CtxPermutedChangedHi}{17.0\xspace}
\newcommand{\CtxPermutedChangedLo}{14.7\xspace}

\newcommand{\DepthBelowRatePerOneZeroZeroZero}{19.4\xspace}

\newcommand{\DepthReferenceExpectedDeepSharePct}{92.6\xspace}

\newcommand{\DepthReferenceObservedDeepSharePct}{95.3\xspace}

\newcommand{\DepthTopRatePerOneZeroZeroZero}{9.8\xspace}

\newcommand{\DepthWithinPairRemovingMeanDiffPctPts}{5.0\xspace}

\newcommand{\DepthWithinPairRemovingReposDeepHigher}{25\xspace}

\newcommand{\DepthWithinPairRemovingReposShallowHigher}{10\xspace}

\newcommand{\DocFormListNamedPct}{12.7\xspace}
\newcommand{\DocFormListUnits}{284\xspace}

\newcommand{\DocFormProseNamedPct}{7.9\xspace}
\newcommand{\DocFormProseUnits}{593\xspace}

\newcommand{\DocLabelNoRemovalPct}{19.1\xspace}

\newcommand{\DocMatchAgree}{46\xspace}
\newcommand{\DocMatchKappa}{0.84\xspace}

\newcommand{\DocMatchRefers}{24\xspace}

\newcommand{\DocMatchSample}{50\xspace}

\newcommand{\DocPairs}{195\xspace}
\newcommand{\DocPairsLabelPct}{67.3\xspace}

\newcommand{\DocPairsNeitherPct}{26.8\xspace}
\newcommand{\DocPairsNoRemoval}{1,043\xspace}

\newcommand{\DocRenameCandPct}{22.3\xspace}
\newcommand{\DocRepos}{36\xspace}

\newcommand{\DocUnits}{1,211\xspace}
\newcommand{\DocUnitsBelowNamedPct}{7.8\xspace}
\newcommand{\DocUnitsBelowTopPct}{86.4\xspace}
\newcommand{\DocUnitsBelowWithNotes}{929\xspace}

\newcommand{\DocUnitsNamedAdjPct}{4.3\xspace}
\newcommand{\DocUnitsNamedMedianRepo}{9.7\xspace}

\newcommand{\DocUnitsNamedPct}{8.9\xspace}
\newcommand{\DocUnitsNamedPctHi}{14.0\xspace}
\newcommand{\DocUnitsNamedPctLo}{5.5\xspace}
\newcommand{\DocUnitsNamedWithNotesPct}{10.0\xspace}

\newcommand{\DocUnitsTopNamedPct}{24.5\xspace}
\newcommand{\DocUnitsTopWithNotes}{147\xspace}

\newcommand{\ESPnetTwoTags}{38\xspace}
\newcommand{\ESPnetTwoTopFirst}{16\xspace}
\newcommand{\ESPnetTwoTopLast}{42\xspace}
\newcommand{\ESPnetTwoTopUnchPct}{54.1\xspace}
\newcommand{\EtaGrowth}{0.07\xspace}
\newcommand{\EtaGrowthP}{0.55\xspace}

\newcommand{\EtaSize}{0.17\xspace}
\newcommand{\EtaSizeP}{0.08\xspace}
\newcommand{\ExtOddsRatio}{8.7\xspace}
\newcommand{\ExtOddsRatioBandHi}{6.6\xspace}
\newcommand{\ExtOddsRatioBandLo}{4.3\xspace}
\newcommand{\ExtOddsRatioHi}{12.5\xspace}
\newcommand{\ExtOddsRatioLo}{6.2\xspace}
\newcommand{\ExtOddsRatioMH}{4.9\xspace}

\newcommand{\FirstGitHubGenDate}{2021-10-26\xspace}

\newcommand{\FunnelBelowPrefilterJudged}{8\xspace}

\newcommand{\FunnelDomainFail}{76\xspace}
\newcommand{\FunnelExcluded}{20\xspace}
\newcommand{\FunnelJudged}{76\xspace}
\newcommand{\FunnelJudgedFromSearch}{46\xspace}
\newcommand{\FunnelPrefilter}{122\xspace}

\newcommand{\FunnelSeedsJudged}{22\xspace}

\newcommand{\HeadingRuleN}{310\xspace}
\newcommand{\HeadingRuleRepos}{13\xspace}
\newcommand{\HeadingVoteAgreePct}{89.0\xspace}

\newcommand{\KappaCodeHi}{0.78\xspace}

\newcommand{\KappaCodeLo}{0.69\xspace}

\newcommand{\KappaRatHi}{0.58\xspace}

\newcommand{\KappaRatLo}{0.19\xspace}

\newcommand{\LabVoteArchList}{1.9\xspace}
\newcommand{\LabVoteArchProse}{7.1\xspace}

\newcommand{\LabVoteInternalList}{33.3\xspace}

\newcommand{\LabVoteNewCapProse}{36.5\xspace}
\newcommand{\LabVoteNoneAllPct}{2.4\xspace}

\newcommand{\LabVoteRatList}{0.9\xspace}
\newcommand{\LabVoteRatProse}{4.3\xspace}

\newcommand{\LifeActive}{9\xspace}

\newcommand{\LifeStable}{33\xspace}
\newcommand{\ListPooledPct}{53.3\xspace}

\newcommand{\ListPooledPctNarrow}{45.5\xspace}

\newcommand{\ListShareMedianRepo}{6.0\xspace}

\newcommand{\LooLooWithNotesMaxPct}{11.2\xspace}
\newcommand{\LooLooWithNotesMinPct}{8.8\xspace}

\newcommand{\LookFixedCoveredWThreePct}{8.8\xspace}

\newcommand{\LookFixedCoveredWZeroPct}{8.5\xspace}

\newcommand{\LookUnitsWithThreePreceding}{1,096\xspace}
\newcommand{\ModGrowthMedian}{1.7\xspace}

\newcommand{\ModGrowthMedianHi}{2.4\xspace}
\newcommand{\ModGrowthMedianLo}{1.5\xspace}

\newcommand{\ModOtherChangedPct}{30.7\xspace}
\newcommand{\ModPairsUnchangedPct}{61.4\xspace}
\newcommand{\ModReposCycleLast}{26\xspace}

\newcommand{\NAdded}{39\xspace}
\newcommand{\NArchived}{5\xspace}

\newcommand{\NCandSearch}{981\xspace}
\newcommand{\NCommitsNonMerge}{114,446\xspace}

\newcommand{\NDropDuplicate}{1,797\xspace}
\newcommand{\NDropHeading}{636\xspace}
\newcommand{\NDropNotrelease}{1,748\xspace}
\newcommand{\NDropOtherversion}{124\xspace}
\newcommand{\NEntries}{20,864\xspace}
\newcommand{\NEntriesRaw}{25,169\xspace}

\newcommand{\NReleasesHarvested}{1,524\xspace}
\newcommand{\NRepos}{56\xspace}

\newcommand{\NReposAnyList}{33\xspace}

\newcommand{\NReposAnyListNarrow}{24\xspace}

\newcommand{\NReposListBeforeGitHub}{10\xspace}
\newcommand{\NReposMajorityList}{14\xspace}

\newcommand{\NReposWithEntries}{54\xspace}
\newcommand{\NReposZeroLast}{19\xspace}
\newcommand{\NSeedsNotInSearch}{11\xspace}
\newcommand{\NSquashRepos}{5\xspace}

\newcommand{\NonActiveShareAll}{83.9\xspace}

\newcommand{\OrigBN}{5\xspace}

\newcommand{\OrigOtherN}{12\xspace}
\newcommand{\OrigOtherNonActive}{10\xspace}
\newcommand{\OtherChangesPct}{51.3\xspace}
\newcommand{\OtherPooledHi}{13.5\xspace}
\newcommand{\OtherPooledLo}{7.5\xspace}

\newcommand{\PkgAllLineagesPairs}{1,369\xspace}
\newcommand{\PkgAllLineagesUnchangedPct}{85.7\xspace}
\newcommand{\PkgTransitions}{2\xspace}

\newcommand{\RareCohortKOneZeroNoRemovalPct}{28.7\xspace}
\newcommand{\RareCohortKOneZeroPairedDiffHi}{27.9\xspace}
\newcommand{\RareCohortKOneZeroPairedDiffLo}{2.7\xspace}
\newcommand{\RareCohortKOneZeroPairedDiffPts}{16.5\xspace}
\newcommand{\RareCohortKOneZeroPairedRepos}{21\xspace}

\newcommand{\RareCohortKOneZeroReleases}{377\xspace}
\newcommand{\RareCohortKOneZeroRemovalPct}{49.7\xspace}

\newcommand{\RareCohortKOneZeroReposPositive}{17\xspace}

\newcommand{\ReadingAnnouncedOfLexicalPct}{50.9\xspace}
\newcommand{\ReadingAnnouncedPct}{30.1\xspace}

\newcommand{\ReadingAnnouncedUnmatchedPct}{28.1\xspace}
\newcommand{\ReadingLexical}{108\xspace}

\newcommand{\ReadingUnmatched}{1,103\xspace}

\newcommand{\RelAfterMixed}{207\xspace}
\newcommand{\RelAfterN}{572\xspace}

\newcommand{\RelEntriesCoded}{448\xspace}
\newcommand{\RelEntryFlaggedHi}{42\xspace}
\newcommand{\RelEntryFlaggedLo}{41\xspace}
\newcommand{\RelEntryKappaCoders}{0.69\xspace}
\newcommand{\RelEntryKappaModelHi}{0.70\xspace}
\newcommand{\RelEntryKappaModelLo}{0.63\xspace}
\newcommand{\RelEntryVoteArch}{31\xspace}

\newcommand{\RelModuleRestrOnlyKappaHi}{0.74\xspace}
\newcommand{\RelModuleRestrOnlyKappaLo}{0.63\xspace}
\newcommand{\RelModuleRestrOnlyYesModel}{5\xspace}
\newcommand{\RelModuleYesCoderHi}{8\xspace}
\newcommand{\RelModuleYesCoderLo}{4\xspace}
\newcommand{\RelModuleYesModel}{14\xspace}

\newcommand{\RelNRemoving}{25\xspace}

\newcommand{\RelReasonN}{50\xspace}
\newcommand{\RelReasonYesCoderHi}{16\xspace}
\newcommand{\RelReasonYesCoderLo}{8\xspace}

\newcommand{\RelSheetN}{50\xspace}

\newcommand{\ReposCycleLast}{25\xspace}
\newcommand{\SizeMax}{75\xspace}
\newcommand{\SizeMedian}{12\xspace}

\newcommand{\SizeMin}{3\xspace}
\newcommand{\SizeQThree}{16\xspace}
\newcommand{\SnapshotDate}{2026-08-15 to 2026-09-27\xspace}

\newcommand{\StructMajorRateRatio}{5.2\xspace}
\newcommand{\StructMajorRateRatioHi}{7.4\xspace}
\newcommand{\StructMajorRateRatioLo}{3.6\xspace}

\newcommand{\StructNExcluded}{4\xspace}
\newcommand{\StructNRepos}{49\xspace}
\newcommand{\StructNReposPkgChange}{7\xspace}
\newcommand{\StructNTags}{1,395\xspace}
\newcommand{\StructNoPkgTags}{33\xspace}
\newcommand{\StructOtherMinorChangedPct}{8.9\xspace}

\newcommand{\StructPairs}{1,346\xspace}

\newcommand{\StructPairsUnchangedPct}{86.1\xspace}

\newcommand{\SymMovedPct}{33.9\xspace}

\newcommand{\SymNamedMedianPct}{3.1\xspace}
\newcommand{\SymNamedPct}{7.4\xspace}
\newcommand{\SymParseErrorFiles}{510\xspace}
\newcommand{\SymParseErrorRepos}{6\xspace}
\newcommand{\SymParseErrorTags}{92\xspace}

\newcommand{\SymReleasesWithRemovals}{351\xspace}

\newcommand{\SymRemovedUnits}{2,926\xspace}
\newcommand{\SymReposWithRemovals}{42\xspace}

\newcommand{\TagsAll}{1,749\xspace}
\newcommand{\TagsRelease}{1,428\xspace}

\newcommand{\TopEntryPct}{21.4\xspace}
\newcommand{\TopEntryRepo}{NeMo\xspace}
\newcommand{\TopGrowthMedian}{1.2\xspace}
\newcommand{\TriageChangesPct}{48.7\xspace}
\newcommand{\TriageHitPct}{33.6\xspace}
\newcommand{\TriagePairsPct}{20.1\xspace}

\newcommand{\UnchangedMedianAll}{85.7\xspace}
\newcommand{\UnchangedMedianAllHi}{89.8\xspace}
\newcommand{\UnchangedMedianAllLo}{80.0\xspace}

\newcommand{\UnparsedTotal}{0\xspace}

\newcommand{\VoteStructOutsideHeadingPct}{5.1\xspace}

%% file: tables/numbers_control.tex
\newcommand{\CtlListShareMedianRepo}{23.6\xspace}

\newcommand{\CtlModGrowthMedian}{5.1\xspace}

\newcommand{\CtlNRepos}{13\xspace}

\newcommand{\CtlNonActiveShareAll}{76.9\xspace}

\newcommand{\CtlSizeMedian}{22\xspace}

\newcommand{\CtlStructPairsUnchangedPct}{82.5\xspace}

%% file: sections/abstract.tex
\begin{abstract}
Developers who upgrade a Python audio library need to know what changed in the code they import and whether the release notes say so. Release notes and breaking changes have been mined for general software, but the module structure of audio libraries and its documentation have not been studied together. We mine \NRepos audio repositories used in machine learning workflows: the import graph of \StructNRepos packages at \StructNTags releases, joined to \NEntries release-note and changelog entries, labeled by three open-weight language models and checked against two human coders. We find that (1) the set of a package's top-level modules is unchanged in \StructPairsUnchangedPct\% of consecutive release pairs, while the set of module files is unchanged in \ModPairsUnchangedPct\%. (2) A major increment changes the top level \StructMajorRateRatio times as often as another increment, and major and 0.x minor increments together account for \TriageChangesPct\% of the observed changes. (3) When a release removes modules, its notes name \DocUnitsNamedPct\% of the removal units under a lexical match, an estimated \DocUnitsNamedAdjPct\% after a hand check of the matches. (4) Longer notes are more likely to contain an entry labeled as structural, and within each length band this proportion is also higher for releases with module removals. We intend to release the aligned dataset and the pipeline.
\end{abstract}

%% file: sections/introduction.tex
\section{Introduction}\label{sec:intro}

Open-source audio machine learning (audio-ML) frameworks are both research code and production dependencies. Libraries such as librosa compute features, toolkits such as ESPnet and SpeechBrain package training pipelines, model packages such as Whisper distribute a published model, and symbolic music libraries such as music21 represent scores. A developer who depends on one of these frameworks has to decide when to upgrade. When Coqui TTS released version 0.0.11, for example, it moved \texttt{tts.layers.gst\_layers} and other modules of \texttt{tts.layers} into new subpackages without keeping the old import paths, and none of the eight entries of its release notes mentions the move. An import of one of these modules fails after the upgrade unless the project kept the old path as an alias. The cost of an upgrade depends on how much of the code the developer imports has changed and on whether the release notes say so.

Semantic versioning reserves a major version for incompatible changes~\cite{raemaekers2017semver,decan2021semver}, and release notes are where a maintainer says what changed. Studies of release notes classify what the notes say~\cite{bi2022releasenotes,wu2022releasenoteissues}. Omissions have been measured against an external list of expected changes: Abebe et al.\ find that most notes list 6 to 26\% of the issues a release addressed~\cite{abebe2016releasenotes}, and Du and Ma find 114 of 405 API breaking changes of 45 Python packages missing from the changelogs~\cite{du2022aexpy}. Studies of architectural evolution measure the code~\cite{behnamghader2017largescale,genfer2024exploring} and leave the documentation aside. We take the module paths and public names that disappear between two releases as the expected changes and ask whether the release's own notes mention them. Our literature search (Section~\ref{sec:related}) found no study that does this for audio libraries or covers their module structure.

We examine version numbers and release notes on \NRepos audio-ML repositories in five archetypes, selected by a written protocol, with \CtlNRepos machine learning libraries from other domains as a comparison cohort. For each package we extract the import graph at every release and join the release to its own release-note and changelog entries, \NEntries in all. We call the modules and subpackages directly inside a package its \emph{top level}, the part a user imports first. We do not equate it with the public API, which projects declare in different ways. Three open-weight language models label every entry for the kind of change it announces. We check the labels against the maintainers' own section headings, against the code history, and against two authors who coded whole releases, and we mark the codes the coders confirm and report the others as model assignments.

We find that structural change happens below the top level of a package, and that the release notes rarely name it. The top level stays the same in \StructPairsUnchangedPct\% of consecutive release pairs. A major increment changes it \StructMajorRateRatio times as often as another increment, and a minor increment of a package still at version 0.x, for which semantic versioning promises no stability, changes it at a rate between the two. Major and 0.x minor increments are \TriagePairsPct\% of the pairs and account for \TriageChangesPct\% of the top-level changes, and the other increments account for the rest. The module files below the top level change in many more releases. When a release removes a module or a public name, its notes name it in fewer than one case in ten under a lexical match. Longer notes carry a structural entry more often, whether or not anything was removed, and within every length band the releases with module removals carry one more often than the others. Lexical naming is uncommon in both forms of notes, lists of pull-request titles and prose.

For a developer, disappearing import paths are candidates for review at an upgrade, with more attention at the increments with the highest change rates. Path stability alone does not establish API or behavioral compatibility. Release-note generators can list the module paths and names a release removed, which the code history records. We organize the study around four research questions:
\begin{itemize}
\item[\textbf{RQ1}] How are audio-ML packages structured at the module level, and does structure differ by archetype?
\item[\textbf{RQ2}] How does their structure change across releases, and what lifecycle do the repositories follow?
\item[\textbf{RQ3}] In which form do maintainers document changes, which kinds of change do they record, and do they state why?
\item[\textbf{RQ4}] When a module path or a public name disappears between two releases, how often do the release's notes mention it?
\end{itemize}
We contribute a release-aligned dataset of import graphs for \StructNRepos Python audio-ML packages at \StructNTags releases, joined to \NEntries labeled entries, which we intend to release with the pipeline. We also contribute a measurement of the documentation gap at the level of modules and of public names, and a code-by-code validation of language-model labels for release notes, which identifies the codes that meet our agreement criterion in the evaluated sample.

Our conclusions concern the selected repositories and the module paths and names extracted from their release histories. We study Python packages on GitHub that pass our selection criteria, and we measure structure at the level of module paths and public names rather than of signatures or behavior. We ask what a release's own notes say, so a removal announced only in an issue, a migration guide, or a deprecation warning in the code counts as unnamed.

%% file: sections/design.tex
\section{Study Design}\label{sec:design}

\input{sections/overview_fig}
Figure~\ref{fig:overview} shows which data sources each research question uses. The mining steps use only the Python standard library and \texttt{git}, and every table, figure, and in-text number is generated from per-item results, which the replication package contains.

\subsection{Corpus and control}\label{sec:corpus}

\paragraph{Scope} We study Python audio libraries used in machine-learning workflows, and we call them audio-ML frameworks for short: the open-source libraries that audio machine learning is built with. The scope includes libraries that ship no model, such as audio I/O, effects, and symbolic music libraries, because ML pipelines depend on them.

\paragraph{Archetypes} The archetype describes what a package is for and who uses it. We assign each package one of five. \emph{A, signal processing} covers signal processing, feature extraction, augmentation, audio I/O, and effects. \emph{B, model-centric package} centers on one model or model family released with a paper. \emph{C, end-to-end toolkit} ships training or pretrained pipelines for several tasks or model families. \emph{D, data and evaluation} loads datasets or computes evaluation metrics and ships no trainable model. \emph{E, symbolic music} reads, writes, or transforms MIDI, MusicXML, or notation and ships no trainable model. When more than one rule matches, the more specific rule wins, in the order E, D, C, B, A.

\paragraph{Selection} We queried the GitHub search API with twelve fixed topic queries (e.g., \texttt{topic:audio}, \texttt{topic:speech-recognition}, \texttt{topic:music-information-retrieval}, each with \texttt{language:Python}), which returned \NCandSearch candidates. The inclusion criteria are: not a fork; at least three years of history or two major versions; at least 300 commits (150 for archetype B, whose packages are small by construction); at least 15 distinct author names; a packaging file at the root and at least one tag; and audio, music, or speech analysis, processing, generation, or learning as the primary domain. Applications (players, downloaders, bots), metadata editors, clients of cloud services, and general ML frameworks with an audio module fail the domain criterion. We added seeds that the topic search misses because of their GitHub topic or language labels, such as pyannote.audio and Whisper. An API prefilter (not a fork, at least 150 commits and 15 contributors, a packaging file, a tag) kept \FunnelPrefilter of the search results. Of these, \FunnelDomainFail failed the domain criterion on their description and \FunnelJudgedFromSearch were judged on the clone, together with \FunnelSeedsJudged seeds that the search does not return and \FunnelBelowPrefilterJudged search results whose API counts fell below the prefilter but whose clone met the thresholds (\FunnelJudged judged candidates). Of these, \FunnelExcluded failed, and the exclusion log records the first failed criterion of each. The remaining \NRepos repositories, with \NCommitsNonMerge non-merge commits in all, entered the corpus (Table~\ref{tab:corpus}), \NAdded more than an initial set of 17 well-known projects. The corpus table in the replication package cites the papers of the corpus systems.

\paragraph{Comparison cohort} To see whether similar patterns appear outside audio, we selected \CtlNRepos ML libraries from vision, text, tabular data, graphs, and time series under the same criteria and archetype rules A to D, chosen for a spread of owners rather than matched or sampled. The cohort goes through every step except labeling.

\paragraph{Owners} Each owner is coded as a company, an academic group, or a community (an independent organization or an individual) from its public profile.

\begin{table}[t]
\caption{Model-centric packages are the smallest and most often company-owned, and most repositories of every archetype are past their peak activity. Columns: repositories, owners (company / academic / community), median commits, authors, releases, lifecycle (active / maintenance / dormant).}
\label{tab:corpus}
\centering\scriptsize\setlength{\tabcolsep}{2.5pt}
\input{tables/corpus_summary.tex}
\end{table}

\subsection{Structure per release (RQ1, RQ2)}\label{sec:structure}

We build the import graph of each repository's main package at every release. A tag is a release when it is a version number without a prerelease marker (\texttt{a}, \texttt{b}, \texttt{rc}, \texttt{alpha}, \texttt{beta}, \texttt{dev}, \texttt{pre}). Of \TagsAll tags, \TagsRelease are releases. Consecutive releases are ordered by version number, because creation order interleaves back-ported releases. A pair is a major increment when the first version component grows. Semantic versioning reserves major increments for incompatible changes and makes no promise while the major version is zero~\cite{raemaekers2017semver,decan2021semver}. Not every project in the corpus follows it, so we read the version numbers as the maintainers wrote them, and the comparison by kind of increment (Table~\ref{tab:toplevel}) shows whether a major number marks larger structural change. We report minor increments at major version zero (such as 0.4 to 0.5) separately and count calendar versions (such as 2024.1) as non-major.

\paragraph{Package} The main package is the import package whose name matches the distribution name declared in \texttt{pyproject.toml}, \texttt{setup.py}, or \texttt{setup.cfg}, looked up at the repository root and under \texttt{src/}, or else the largest top-level Python package that is not a test directory. When the selected package differs between releases, or no package is found at some releases, we use the repository's most frequent package and drop the other releases (\StructNoPkgTags releases in \StructNReposPkgChange repositories, named in the replication package). Three repositories with C or C++ cores and \StructNExcluded whose code is not a regular package (two single-file modules, one namespace package, and one without a detectable package) are excluded from this analysis and kept for the others.

\paragraph{Graphs} At the \emph{top level}, a node is a top-level submodule of the package: a module file or a subpackage. At the \emph{module level}, a node is a module file. An edge $u \rightarrow v$ exists when a file in $u$ imports $v$, including imports inside function bodies. Relative imports resolve against the importing file's directory, and \texttt{from pkg import name} produces an edge only when \texttt{name} is a module present at that release. The graphs are explicit static import relations. We report nodes (S), edges (E), and strongly connected components with more than one node (cycles). A release pair leaves the top level unchanged when the set of submodule names is the same rather than only their number.

\subsection{Lifecycle and change (RQ2)}\label{sec:lifecycle}

We count non-merge commits on the default branch by author date, up to the snapshot, the clone date (\SnapshotDate, the initial 17 repositories first). A repository is \emph{dormant} when it has no commit in the twelve months before the snapshot. It is in \emph{maintenance} when its trailing twelve-month count is below 25\% of its maximum over any twelve consecutive months, and \emph{active} otherwise. Maintenance is relative to a repository's own peak, so it includes steadily developed projects whose activity has declined, such as librosa and SpeechBrain. We combine the maintenance and dormant categories as \emph{not active} for the comparison below. A repository archived on GitHub is dormant. We report the classification under six combinations of window (6, 12, 24 months) and fraction (0.25, 0.5). On the initial 17 repositories, we had observed that model-centric packages are more often not active than the other archetypes pooled, and we test this prediction on the \NAdded repositories added afterwards. The comparison pools archetypes A, C, D, and E.

\subsection{Documented changes (RQ3, RQ4)}\label{sec:docs}

\paragraph{Sources} We harvested all GitHub Releases through the REST API (\NReleasesHarvested releases) and parsed the in-tree changelogs. Each bullet with its continuation lines, or each prose paragraph, is one entry. We then remove four kinds of entry: notes of tags that the release filter rejects, such as prereleases (\NDropNotrelease); sections of a release page that reproduce the changelog of an earlier release (\NDropOtherversion); headings and markup parsed as entries (\NDropHeading); and repeats of the same text within a repository, such as a change published in both the changelog and a GitHub Release or a pull request listed in several patch releases (\NDropDuplicate, keeping the earliest dated copy). This leaves \NEntries of \NEntriesRaw entries over \NReposWithEntries repositories. The duplicate rule keeps one copy of a repeated text. Section~\ref{sec:rq4} checks its effect on the removal result.

\paragraph{Form} Each entry receives one form, whatever tool produced it. A \emph{list} entry is a title followed by an author handle and a pull-request reference, with nothing after the reference; a title followed by a parenthesized pull-request reference alone; or a commit-title line. This rule covers GitHub's release-note generator, NeMo's and ESPnet's own formats, SpeechBrain's commit lists, and the reference-only lists of torchaudio, madmom, Asteroid, and Lhotse. A narrower rule that also requires the handle is reported as a sensitivity (Section~\ref{sec:threats}). GitHub's generator lists merged pull requests and contributors~\cite{github2026releasenotesdocs}. A \emph{contributor} entry is a first-contribution line or a thanks line listing at least three handles. Contributor entries are counted but not labeled. All other entries are \emph{prose}. A sub-flag marks GitHub's generator format, which dates each repository's adoption of that tool.

\paragraph{Change taxonomy} Each entry receives one code from a fixed seven-code taxonomy with a stated precedence: when more than one code fits, the first in the list wins. Three codes describe architecturally relevant announcements: \texttt{BREAKING} (a public function, class, argument, default, format, or module is removed, renamed, or changed so that user code may stop working), \texttt{DEPRECATION} (an element still works but will be removed later), and \texttt{RESTRUCTURING} (a module or component is moved, split, merged, or renamed with compatibility kept). Four describe other changes: \texttt{NEW\_CAPABILITY}, \texttt{BUGFIX}, \texttt{INTERNAL} (build, packaging, tests, dependencies, refactoring, performance, documentation), and \texttt{OTHER}. The labeler also states whether the entry gives a reason for the change (the rationale flag). We call the code that at least two of the three models give an entry its \emph{majority label}, and we call the first three codes the \emph{architectural codes}. A context ablation on the coded releases, with the taxonomy fixed, hides the heading, hides the entry, or shows the heading of another release (Section~\ref{sec:validation}). The labels of an earlier ten-code variant are in the replication package.

\paragraph{Labeling} Three open-weight models from two families label every entry: Qwen2.5-7B, Qwen2.5-32B, and Llama-3.3-70B, served locally with Ollama in 4-bit quantization (Q4\_K\_M), temperature zero, and the model digest recorded. All three receive the same prompt, which shows the repository, the release, the section heading above the entry, and the entry, lists the seven codes with a one-line definition and one example each, and asks for a JSON object. A reply without JSON is accepted when it names exactly one code. Over the full run, \texttt{OTHER} and unparsable outputs together stay below 30\% for every labeler (\OtherPooledLo--\OtherPooledHi\%, \UnparsedTotal unparsable).

\paragraph{Validation of the labels} Every corpus-wide number about the kind of change rests on the model labels, so we check them in three ways before we use them, and the acceptance rule below decides which codes the results rely on (Section~\ref{sec:validation}). First, as a consistency check against the maintainers' headings, which the models also see: for the entries that a maintainer filed under a heading that says breaking, incompatible, deprecated, or removed, we report how often the majority label is architectural. Second, against the code: we report the odds that a release carries an architectural entry when it removes a non-private module path against when it does not. Third, against two authors, who read whole releases rather than isolated entries. Blind to the model labels and to each other, they code \RelSheetN releases, half of which remove a non-private module path, matched on the number of entries, with at most 30 entries each: for each release, whether the notes announce a breaking change, a deprecation, a removed or moved module, and a reason, and which entries do so. The models' answer for a release is yes when the majority label of any of its entries is the matching code. We compare it with each coder's answer question by question, and the flagged entries give a per-entry comparison on the architectural codes. Following Ahmed et al.~\cite{ahmed2025llmannotation}, a model replaces at most one coder. We use the model labels for corpus-wide numbers of a code only if Cohen's $\kappa$ between the models and each coder is at least the coders' mutual $\kappa$ minus 0.10 and at least 0.667, the tentative threshold of the guidelines we follow~\cite{baltes2025llmguidelines}. Otherwise the corpus-wide numbers for that code are reported as exploratory model assignments. The same two coders judge 50 of the name matches of the removal analysis below. Table~\ref{tab:human} gives the outcome.

\paragraph{Removals (RQ4)} For each pair of consecutive parsed releases, we list the non-private module paths of the main package (no path component that starts with \texttt{\_}, no test directory) present at the earlier release and absent at the later one, using only the git trees. The rule is a naming convention rather than an audited interface. A removal is \emph{named} when an entry of the later release's notes contains the module's basename or dotted path as a word, a lexical match. The match is liberal, since a generic name such as \texttt{data} counts, so we also report the share of a sample of 50 matches that refers to the removed module. Files inside the package directory that are not importable as modules, such as example scripts, count as module paths, and an importability proxy (every parent directory a package, no examples or scripts directory in the path) is reported beside the result. A release is \emph{labeled} when the majority label of one of its entries is architectural. Below the module, we list the public names of each non-private module at each release: the module-level functions and classes whose name does not start with an underscore, and the names in \texttt{\_\_all\_\_}. A name present at the earlier release and absent at the later one, while its module remains, is a removed name, and a name that appears in another module of the package at the same release is a move candidate. Files that the Python~3 parser rejects, mostly Python~2 code in old releases, are skipped and counted. The same word match applies to names.

\paragraph{Aggregation} \TopEntryRepo contributes \TopEntryPct\% of all entries, so pooled frequencies would describe one project's release-note style. We report frequencies per repository, and archetype summaries are medians over repositories. Intervals for medians are 95\% percentile bootstrap intervals over repositories (2,000 resamples). The audio corpus and the control are compared per repository with Cliff's $\delta$ as a descriptive effect size. The control was chosen for spread rather than sampled, so we attach no $p$-value to it. All tests are exploratory. The unit of the top-level results is the release pair, pooled and as a median over repositories. The unit of the removal results is the removal unit, pooled, with a leave-one-repository-out range beside the pooled share. The unit of the label results is the entry, aggregated per repository. Five terms recur in the results: list and prose entries (the two forms), the architectural codes, the majority label, a labeled release, and a removal unit (a removed module file, or a removed subpackage counted once).

\paragraph{Predictions} (1) Model-centric packages are more often not active than the other archetypes pooled, as on the initial 17 repositories, and an equal or lower share on the added repositories refutes it. (2) If the labels follow the code, releases that remove a non-private module path carry an architectural entry more often than the others. (3) If the models follow the maintainers, entries under breaking or deprecation headings receive an architectural label far more often than other entries. (4) If the comparison cohort shows the same stable top level, a similar stability occurs in the selected non-audio libraries.

\paragraph{Checks of the instruments} Known-answer cases check the implementation of each rule: nineteen for the import parser, eighteen real tag names for the release filter, thirteen for the changelog formats, and seventeen entries for the form rule. The form rule also reproduces a published observation: Jiang et al.\ report that more than half of the GitHub projects with release notes produce them from pull requests~\cite{jiang2021deeprelease}, and \NReposAnyList of our \NReposWithEntries repositories with notes use the list form at least once. Section~\ref{sec:validation} reports three checks of the labels.

%% file: sections/overview_fig.tex
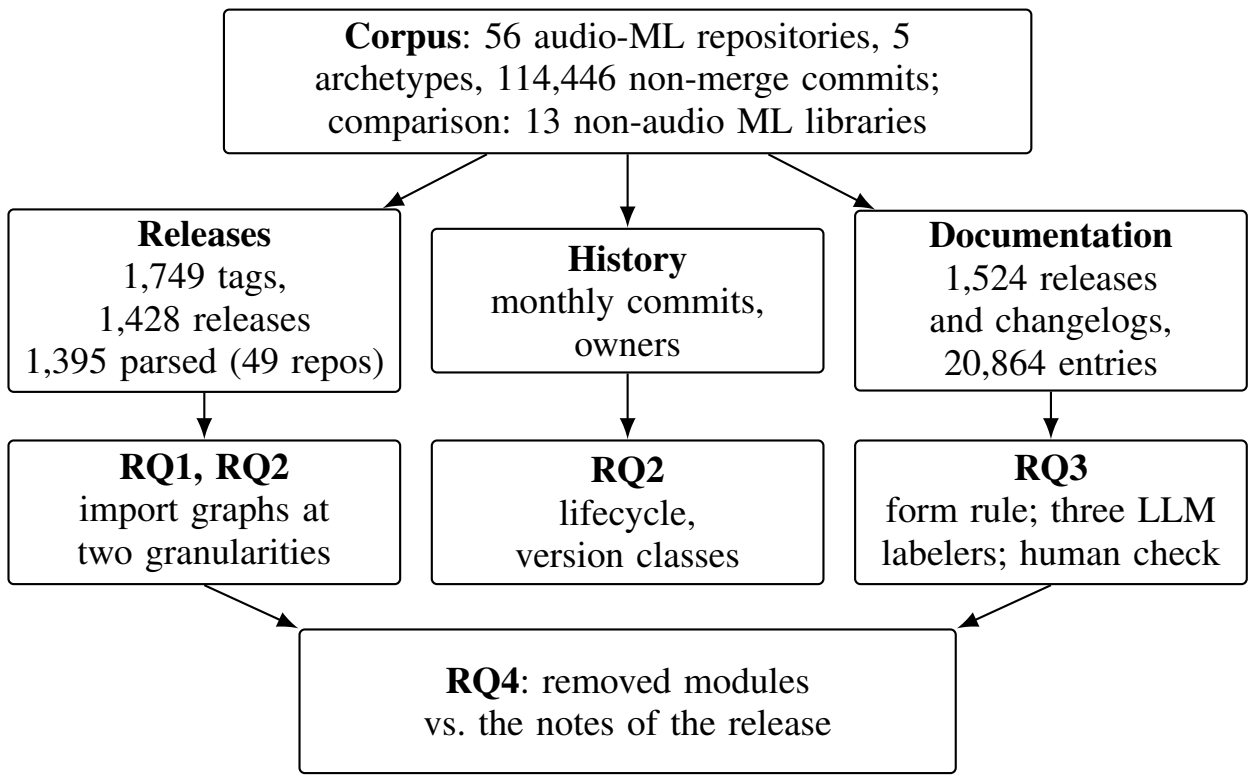
\begin{figure}[t]
\centering
\scriptsize
\begin{tikzpicture}[box/.style={draw, rounded corners=1pt, align=center, inner sep=2pt, text width=2.45cm, minimum height=0.95cm}, >=latex]
\node[box, text width=5.2cm] (corpus) at (2.8,0) {\textbf{Corpus}: \NRepos audio-ML repositories, 5 archetypes, \NCommitsNonMerge non-merge commits; comparison: \CtlNRepos non-audio ML libraries};
\node[box] (tags) at (0,-1.45) {\textbf{Releases}\\ \TagsAll tags, \TagsRelease releases\\ \StructNTags parsed (\StructNRepos repos)};
\node[box] (git) at (2.8,-1.45) {\textbf{History}\\ monthly commits,\\ owners};
\node[box] (notes) at (5.6,-1.45) {\textbf{Documentation}\\ \NReleasesHarvested releases and changelogs, \NEntries entries};
\node[box] (rq1) at (0,-2.85) {\textbf{RQ1, RQ2}\\ import graphs at two granularities};
\node[box] (rq2) at (2.8,-2.85) {\textbf{RQ2}\\ lifecycle, version classes};
\node[box] (rq3) at (5.6,-2.85) {\textbf{RQ3}\\ form rule; three LLM labelers; human check};
\node[box, text width=4.2cm] (rq4) at (2.8,-4.1) {\textbf{RQ4}: removed modules vs.\ the notes of the release};
\draw[->] (corpus) -- (tags); \draw[->] (corpus) -- (git); \draw[->] (corpus) -- (notes);
\draw[->] (tags) -- (rq1); \draw[->] (git) -- (rq2); \draw[->] (notes) -- (rq3);
\draw[->] (rq1.south) -- (rq4.north west); \draw[->] (rq3.south) -- (rq4.north east);
\end{tikzpicture}
\caption{Each research question draws on one or two of the three data sources mined from the corpus.}
\label{fig:overview}
\end{figure}

%% file: tables/corpus_summary.tex
\begin{tabular}{@{}lrcrrrc@{}}
\toprule
Archetype & Repos & Owners & Commits & Authors & Releases & Lifecycle \\
\midrule
A: Signal processing & 16 & 3/4/9 & 990 & 30 & 27 & 2/11/3 \\
B: Model-centric package & 12 & 9/0/3 & 320 & 37 & 13 & 3/5/4 \\
C: End-to-end toolkit & 17 & 7/6/4 & 2,407 & 118 & 17 & 3/9/5 \\
D: Data and evaluation & 5 & 0/1/4 & 840 & 33 & 11 & 0/4/1 \\
E: Symbolic music & 6 & 1/3/2 & 1,365 & 39 & 34 & 1/3/2 \\
\bottomrule
\end{tabular}

%% file: sections/results.tex
\section{Results}\label{sec:results}

\subsection{RQ1: Module structure}\label{sec:rq1}

\begin{table}[t]
\caption{Top levels are small in every archetype, and most release pairs leave them unchanged. Columns: top-level submodules at the last release (S), median share of release pairs with an unchanged top-level set (Unch.), pairs whose count changes by a factor of 1.5 or more (Large), median growth in module files, packages with a top-level / module-level cycle.}
\label{tab:structure}
\centering\scriptsize\setlength{\tabcolsep}{3pt}
\input{tables/structure_summary.tex}
\end{table}

Audio-ML packages present a small top level, and its size varies substantially within archetypes. At the last release, the median package has \SizeMedian top-level submodules, three quarters have at most \SizeQThree, and the range is \SizeMin to \SizeMax (Table~\ref{tab:structure}). On a logarithmic scale, archetype accounts for \EtaSize of the variance in top-level size (permutation test, $p=\EtaSizeP$, exploratory) and \EtaGrowth of the variance in module growth ($p=\EtaGrowthP$).

Import cycles are common: \ReposCycleLast of \StructNRepos packages contain one at the top level and \ModReposCycleLast at module level.

A second-generation package that grows beside the first is one way the top level changes. ESPnet introduced \texttt{espnet2} beside \texttt{espnet}. As its own series it grows from \ESPnetTwoTopFirst to \ESPnetTwoTopLast top-level submodules over \ESPnetTwoTags releases, and only \ESPnetTwoTopUnchPct\% of its release pairs leave the top level unchanged, against \UnchangedMedianAll\% for the median package.

\answer{1}{Audio-ML packages expose a small top level whose sizes overlap across archetypes. Import cycles are common at both granularities.}

\subsection{RQ2: Structural change across releases}\label{sec:rq2}

\begin{figure}[t]
\centering
\includegraphics[width=0.96\columnwidth]{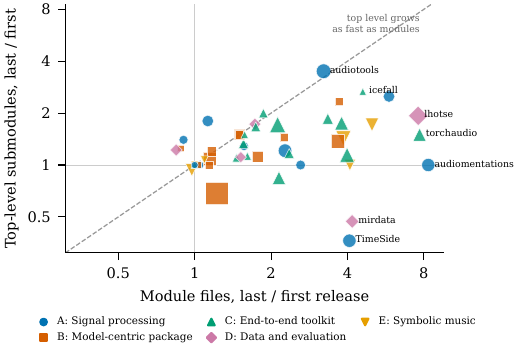}
\caption{Most packages grew more in module files than at the top level. Each marker is a package, placed by its growth in module files (x) and in top-level submodules (y) from its earliest to its latest parsed release, on logarithmic axes. Marker size grows with the number of releases.}
\label{fig:structure}
\end{figure}

\begin{table}[t]
\caption{Major increments change the top level at the highest rate, and the more numerous other increments account for half of the changes. Each row gives the consecutive release pairs of one kind, the pairs that change the set of top-level submodules, the rate, and the share of all top-level changes. Calendar versions count as non-major.}
\label{tab:toplevel}
\centering\scriptsize
\input{tables/toplevel_change.tex}
\end{table}

The top level rarely changes, and the modules below it grow and disappear at a higher rate per path. Of \StructPairs pairs of consecutive releases, \StructPairsUnchangedPct\% leave the set of top-level submodules unchanged (median over repositories \UnchangedMedianAll\%, 95\% interval \UnchangedMedianAllLo--\UnchangedMedianAllHi). The set of module files stays the same in \ModPairsUnchangedPct\% of the pairs. The median package ends with \ModGrowthMedian times (\ModGrowthMedianLo--\ModGrowthMedianHi) as many module files as it started with, against \TopGrowthMedian times as many top-level submodules, and in Figure~\ref{fig:structure} most packages lie below the diagonal. Per eligible path (removed files over the module files present at the earlier release, pooled over consecutive pairs), a module file inside a subpackage disappears at \DepthBelowRatePerOneZeroZeroZero per thousand module-pairs against \DepthTopRatePerOneZeroZeroZero for a file directly in the package, and within pairs with module removals the deeper rate is higher by \DepthWithinPairRemovingMeanDiffPctPts points on average (\DepthWithinPairRemovingReposDeepHigher repositories against \DepthWithinPairRemovingReposShallowHigher). Subpackages contain most module files and account for most removals: \DepthReferenceObservedDeepSharePct\% of the removed files lie in subpackages, against \DepthReferenceExpectedDeepSharePct\% expected if each removing pair's removals fell uniformly on its eligible files.

Top-level change rates differ by kind of increment, and the rate is highest at major increments. Table~\ref{tab:toplevel} splits the release pairs by kind of increment. A major release changes the top level about \StructMajorRateRatio times as often as another release (\StructMajorRateRatioLo--\StructMajorRateRatioHi), and the later majors change it more often than the first. Minor increments of version-zero packages have the next-highest change rate. \NReposZeroLast of the \StructNRepos packages are still at version zero at their last release, and there a minor increment changes the top level at a rate between the majors and the other non-major increments. Major and 0.x minor increments together are \TriagePairsPct\% of the pairs. \TriageHitPct\% of them change the top level, and they account for \TriageChangesPct\% of all top-level changes, so the other increments account for \OtherChangesPct\%. The module files change more often at every kind of increment: \ModOtherChangedPct\% of the other non-major releases change the set of module files, against \StructOtherMinorChangedPct\% that change the top level.

Most repositories are past their peak activity. At the snapshot, \LifeActive repositories are active and the others are in maintenance or dormant, \NArchived of them archived on GitHub (Table~\ref{tab:corpus}), and \LifeStable of \NRepos keep their class under all six threshold combinations. On the initial 17 repositories, all \OrigBN model-centric packages were not active, against \OrigOtherNonActive of \OrigOtherN others. On the \NAdded repositories added afterwards, prediction 1 is refuted: \AddedBNonActive of \AddedBN model-centric packages are not active, against \AddedOtherNonActive of \AddedOtherN others, also when archetype E is left out of the pool. Dormancy is more common among company-owned repositories (odds ratio \AllORCompanyDormant over the corpus), with every interval including one.

\answer{2}{Structural change is more frequent below the top level. The top level changes at a higher rate at major and 0.x minor increments, which account for about half of the changes. The predicted higher inactivity of model-centric packages did not hold in the added repositories.}

\subsection{Validation of the model labels}\label{sec:validation}

\begin{table}[t]
\caption{The models match the coders on breaking changes and deprecations and not on moved modules and reasons, at release level, on releases with at most 30 entries. Two authors (A, B) answered each question for \RelSheetN whole releases. The models (M) answer yes when the majority label of any entry is the matching code. Columns give the releases both coders answered ($n$), the yes counts, Cohen's $\kappa$ for each pair, and whether the models are accepted.}
\label{tab:human}
\centering\scriptsize\setlength{\tabcolsep}{2.2pt}
\input{tables/human_validation.tex}
\end{table}

We compare the model assignments with the maintainers' headings, with the code history, and with the human annotations, and we distinguish validated release-level results from exploratory summaries of model assignments. The models pass the acceptance rule for breaking changes and deprecations and fail it for moved modules and reasons.

We first compare the model assignments with the maintainers' headings and with the observed code changes. On the headings, \HeadingRuleRepos repositories file \HeadingRuleN entries under a heading that says breaking, incompatible, deprecated, or removed, and the majority label is architectural for \HeadingVoteAgreePct\% of them, against \VoteStructOutsideHeadingPct\% of the entries outside such headings. Prediction 3 holds as a consistency check, since the models see the headings. A context ablation asks what the heading contributes: on the \CtxEntries labeled entries of the coded releases the models relabel each entry with the heading hidden, with the entry hidden, and with the heading of another release. The full context reproduces the main run for \CtxFullReproLo--\CtxFullReproHi\% of the entries. Hiding the heading changes \CtxBodyChangedLo--\CtxBodyChangedHi\% of a model's codes and leaves $\kappa$ with the coders where it was (\CtxBodyBreakKappaLo--\CtxBodyBreakKappaHi on breaking changes, \CtxBodyEntryKappaLo--\CtxBodyEntryKappaHi per entry). The heading alone changes \CtxHeadingChangedLo--\CtxHeadingChangedHi\% and drops $\kappa$ to \CtxHeadingBreakKappaLo--\CtxHeadingBreakKappaHi on breaking changes and \CtxHeadingEntryKappaLo--\CtxHeadingEntryKappaHi per entry, and a heading from another release changes \CtxPermutedChangedLo--\CtxPermutedChangedHi\%. The labels rest mainly on the entry text.

The human coders confirm the model labels for breaking changes and deprecations and do not confirm them for moved modules and reasons. Two authors read \RelSheetN whole releases with \RelEntriesCoded entries, \RelNRemoving of which remove a non-private module path, blind to the labels and to each other. For each release they answered whether its notes announce a breaking change, a deprecation, a removed or moved module, and a reason. Table~\ref{tab:human} compares their answers with the models' answers, question by question. On breaking changes and on deprecations, the models agree with each coder at least as well as the coders agree with each other, which supports their release-level use for these two codes within the validation sample.

On a removed or moved module, the coders agree with each other less, and the models say yes for \RelModuleYesModel releases where the coders say yes for \RelModuleYesCoderLo and \RelModuleYesCoderHi. Most of this surplus comes from \texttt{BREAKING} entries that remove a function or a method, drop support for a dependency, or change a format, and name no module. When the question is compared with \texttt{RESTRUCTURING} alone, the models say yes for \RelModuleRestrOnlyYesModel releases and reach $\kappa$ of \RelModuleRestrOnlyKappaLo--\RelModuleRestrOnlyKappaHi, with the agreement for one coder below the threshold. On whether a reason is given, the two coders disagree with each other (\RelReasonYesCoderLo and \RelReasonYesCoderHi of \RelReasonN releases). Agreement between the models and the coders is similarly limited.

Per entry, the coders flag \RelEntryFlaggedLo--\RelEntryFlaggedHi of the \RelEntriesCoded entries as architectural and the majority label \RelEntryVoteArch, with $\kappa$ of \RelEntryKappaCoders between the coders and \RelEntryKappaModelLo--\RelEntryKappaModelHi between the models and a coder.

For the rest of the paper, the shares of breaking changes and deprecations rest on labels validated at release level, and the entry-level shares in Table~\ref{tab:labels} are model-assigned shares with the two validated codes marked. The other codes and the stated reasons describe what the models find. The main measure of RQ4, whether the notes name a removed module or name, uses no model label. It rests on a word match and on the coders' check of the matches.

\subsection{RQ3: Forms and content of release notes}\label{sec:rq3}

\begin{table}[t]
\caption{List entries concentrate in toolkits and in data and evaluation libraries. Columns give the median share of list and contributor entries (List), the repositories with any / a majority of list entries, the repositories that use GitHub's generator format, and the median share of prose entries that reference a pull request or issue.}
\label{tab:genre}
\centering\scriptsize\setlength{\tabcolsep}{3pt}
\input{tables/genre_summary.tex}
\end{table}

Release notes come in two forms, and a minority of repositories writes most of the lists. Release notes or a changelog exist for \NReposWithEntries of \NRepos repositories. A list entry reads ``fixing the convtasnet causal=True bug by @ycemsubakan in https://github.com/.../pull/1851'' (SpeechBrain), and a prose entry reads ``From asteroid 0.5.0 onwards, only torch>=1.8.0 will be supported'' (Asteroid v0.4.5).

Pooled over all \NEntries entries, \ListPooledPct\% are list or contributor entries. Per repository, \NReposAnyList repositories use the list form at least once, \NReposMajorityList write mostly lists, and the median repository writes \ListShareMedianRepo\% of its entries as lists (Table~\ref{tab:genre}). A narrower rule that requires an author handle gives \ListPooledPctNarrow\% pooled and \NReposAnyListNarrow repositories with any list.

The list form predates GitHub's generator (\NReposListBeforeGitHub repositories wrote lists before its first observed use in \FirstGitHubGenDate) and is often partial within a repository: of the \RelAfterN releases after a repository's first list release, \RelAfterMixed mix both forms.

\begin{table}[t]
\caption{Model-assigned shares of the codes, by form. Each cell is the median over repositories of the share of entries (\%) whose majority label, the code at least two of the three models give, is that code. Codes marked $\dagger$ passed the release-level validation (Table~\ref{tab:human}); the other rows describe the models. Repositories with fewer than ten entries of a form are omitted from that column.}
\label{tab:labels}
\centering\scriptsize
\input{tables/labels_vote.tex}
\end{table}

In the model-assigned shares, \texttt{NEW\_CAPABILITY} is the largest code for prose and \texttt{INTERNAL} for lists. Table~\ref{tab:labels} gives, for each code, the share of entries in the median repository whose majority label is that code. In prose, \texttt{NEW\_CAPABILITY} leads with \LabVoteNewCapProse\%. In lists, \texttt{INTERNAL}, the code of build, dependency, and refactoring titles, leads with \LabVoteInternalList\%. The three architectural codes together cover \LabVoteArchProse\% of prose and \LabVoteArchList\% of list entries in the median repository.

All three models give the same code to \AgreeAllThreePct\% of entries, their pairwise Cohen's $\kappa$ lies between \KappaCodeLo and \KappaCodeHi, and \LabVoteNoneAllPct\% of entries have no majority label. The models flag a reason in \LabVoteRatProse\% of prose and \LabVoteRatList\% of list entries in the median repository, but the rationale judgments did not pass the validation, so the prevalence of stated reasons remains uncertain under our instrument.

\answer{3}{Release notes come in two forms, and a minority of repositories writes most of the lists. In the model assignments, breaking changes and deprecations are a small share of the entries in either form. The prevalence of stated reasons remains uncertain under our instrument.}

\subsection{RQ4: Documentation of removals}\label{sec:rq4}

\begin{table}[t]
\caption{Few removed modules are named in the notes of the release that removes them. The rows follow the removals from release pairs to the names in the notes. A removal unit is a removed module file or a removed subpackage counted once.}
\label{tab:funnel}
\centering\scriptsize
\input{tables/rq4_funnel.tex}
\end{table}

Under the lexical match, release notes name fewer than one removed module path in ten. Table~\ref{tab:funnel} follows the removals from release pairs to the notes. In \DocPairs of the \StructPairs release pairs, in \DocRepos repositories, the later release removes non-private module paths, and these removals form \DocUnits removal units once a removed subpackage is counted once. Like the growth, most removals are below the top level (\DocUnitsBelowTopPct\% of the units). The notes name \DocUnitsNamedPct\% of the units (95\% interval \DocUnitsNamedPctLo--\DocUnitsNamedPctHi), counting removals in releases without notes as unnamed, and \DocUnitsNamedWithNotesPct\% of the units in releases with notes. The median repository names \DocUnitsNamedMedianRepo\%. Restoring to their releases the repeated entries that the duplicate rule removed names \AuditDedupRestoredUnits more unit (\AuditDedupNamedPct\%). Dropping the one pair with module removals that spans a skipped release gives \AuditAdjacentNamedPct\%. Leaving out any one repository moves the share between \AuditLooNamedPctMin\% and \AuditLooNamedPctMax\% (\LooLooWithNotesMinPct--\LooLooWithNotesMaxPct\% with notes), although NeMo contributes \AuditUnitsTopRepoShare\% of the units.

The name match is liberal, since a generic name such as \texttt{data} counts. Two authors judged \DocMatchSample matched names and agree on \DocMatchAgree ($\kappa=\DocMatchKappa$). Both find that \DocMatchRefers of the \DocMatchSample refer to the removed module. The rest are a generic word, a URL, or another module with the same file name, which gives an estimated \DocUnitsNamedAdjPct\% of the units with a match that refers to the removed module. The estimate multiplies the lexical rate by the audited precision and leaves unmeasured the removals that the notes announce in other words. Counting a deprecation word in the three preceding releases raises the named share of the \LookUnitsWithThreePreceding units that have them from \LookFixedCoveredWZeroPct\% to \LookFixedCoveredWThreePct\%. Coqui TTS v0.0.11, for example, moves \texttt{tts.layers.gst\_layers} and other modules of \texttt{tts.layers} into new subpackages without keeping the old import paths, and none of the release's eight entries mentions the move. Of the removed files, \DocRenameCandPct\% reappear under the same name elsewhere in the package, which marks them as move candidates. A move can still break the old import path. A model reading explores what the lexical match misses. The three labeling models read every unit against the notes of its release and answered whether the notes announce the removal or move in any words. The majority of the three says yes for \ReadingAnnouncedPct\% of the units: for \ReadingAnnouncedOfLexicalPct\% of the \ReadingLexical units with a lexical match and for \ReadingAnnouncedUnmatchedPct\% of the \ReadingUnmatched without one, such as ``complete refactor of libroom into C++'' for the removed \texttt{c\_package} of pyroomacoustics. This exploratory result suggests that the lexical match misses announcements written in other words. The announcement judgments are not validated against human annotations: on the \DocMatchSample matched names the coders judged, the majority agrees with both coders on 46 about the referent, which says nothing about the unmatched units. Two rule-based proxies point the same way: among the \AuditUnnamedUnits unnamed units, the notes mention a parent package of the module for \AuditUnnamedParentMentioned and contain a removal word in an entry that names a path component for \AuditUnnamedRemovalWordComponent. By the importability proxy, \AuditImportableUnitsPct\% of the units are importable module paths. The notes name \AuditImportableNamedPct\% of them against \AuditNonImportableNamedPct\% of the others.

Lexical mentions are also infrequent for names removed from modules that remain. Across \SymReleasesWithRemovals release pairs in \SymReposWithRemovals repositories, \SymRemovedUnits public names disappear while their module remains, and \SymMovedPct\% of them reappear in another module of the package at the same release. Of the names removed in releases with notes, the notes mention \SymNamedPct\%. The median repository mentions \SymNamedMedianPct\%.

\begin{table}[t]
\caption{Longer notes carry an architectural entry more often, and within every length band the releases with module removals carry one more often than the others. Each cell gives the share of releases with an entry whose majority label is architectural, by length of the notes, with the within-band odds ratio.}
\label{tab:rq4bins}
\centering\scriptsize
\input{tables/rq4_label_bins.tex}
\end{table}

Longer notes carry a structural entry more often, and so do releases with module removals within every length band. The entry here is any entry whose majority label is one of the three architectural codes, a model assignment that includes the unvalidated \texttt{RESTRUCTURING}. Among releases with notes, \DocPairsLabelPct\% of those with a module removal carry an entry whose majority label is architectural, against \DocLabelNoRemovalPct\% of the \DocPairsNoRemoval without removals, a crude odds ratio of \ExtOddsRatio (\ExtOddsRatioLo--\ExtOddsRatioHi). Prediction 2 holds: releases with module removals are more likely to contain an entry assigned an architectural label. Part of this gap comes from length. Removing releases have longer notes, and a release counts as labeled when any one entry carries the code, so a longer note has more chances to count. Table~\ref{tab:rq4bins} stratifies by length. Within each band the releases with module removals carry an architectural entry more often, with odds ratios between \ExtOddsRatioBandLo and \ExtOddsRatioBandHi and a Mantel-Haenszel summary of \ExtOddsRatioMH. The difference also survives an equal reading budget. Among the \RareCohortKOneZeroReleases releases with at least ten entries, the exact chance that ten entries drawn at random from a release include an architectural one, averaged over releases, is \RareCohortKOneZeroRemovalPct\% for the releases with module removals against \RareCohortKOneZeroNoRemovalPct\% for the others, a within-repository difference of \RareCohortKOneZeroPairedDiffPts points (the mean over repositories with both kinds of release, bootstrap over repositories \RareCohortKOneZeroPairedDiffLo--\RareCohortKOneZeroPairedDiffHi, positive in \RareCohortKOneZeroReposPositive of \RareCohortKOneZeroPairedRepos repositories). These are two associations, and we do not rank the importance of length and removal. Among the releases with removals and notes, \DocPairsNeitherPct\% neither name a removed module nor carry an architectural entry.

The notes name a larger share of the top-level removals, the ones a user imports first. In releases with notes, they name \DocUnitsTopNamedPct\% of the \DocUnitsTopWithNotes removal units at the top level and \DocUnitsBelowNamedPct\% of the \DocUnitsBelowWithNotes below it, by the liberal match, which overstates both shares. The notes name top-level removals about three times as often as deeper ones, and even at the top level three in four go unnamed under the lexical match.

Lexical naming is uncommon in both forms of notes. By the liberal match, releases whose notes are mostly lists name \DocFormListNamedPct\% of their \DocFormListUnits removal units, and releases written only in prose name \DocFormProseNamedPct\% of \DocFormProseUnits. The comparison is unadjusted.

\answer{4}{Under a lexical match, release notes rarely name what a release removes, for module paths and public names alike, and least often below the top level. Longer notes and releases with module removals both carry structural entries more often. Identifying the affected paths requires comparing the code across releases.}

\subsection{The non-audio comparison cohort}\label{sec:control}

The comparison cohort of \CtlNRepos libraries from other domains also keeps its top level in most release pairs, as prediction 4 states: \CtlStructPairsUnchangedPct\% of its release pairs leave the top level unchanged, and the per-repository shares are alike (Cliff's $\delta=\CliffUnchanged$). The audio packages are smaller (median \SizeMedian against \CtlSizeMedian top-level submodules, $\delta=\CliffSize$) and grow more slowly at module level (\ModGrowthMedian against \CtlModGrowthMedian times, $\delta=\CliffGrowth$). The median cohort repository writes \CtlListShareMedianRepo\% of its entries as lists and the median audio repository \ListShareMedianRepo\% ($\delta=\CliffList$), and the shares of repositories that are not active are similar (\NonActiveShareAll\% against \CtlNonActiveShareAll\%). The cohort's per-archetype table is in the replication package.

%% file: tables/structure_summary.tex
\begin{tabular}{@{}lrrcrrrc@{}}
\toprule
Arch. & Repos & Rel. & Top-level S & Unch.\,\% & Large & Growth & Cycles \\
\midrule
A & 11 & 323 & 3--27 & 92.2 & 3 & 2.3 & 7/7 \\
B & 12 & 391 & 6--25 & 86.4 & 3 & 1.4 & 2/3 \\
C & 15 & 396 & 6--24 & 83.3 & 4 & 2.1 & 11/11 \\
D & 5 & 94 & 8--27 & 75.0 & 1 & 1.7 & 1/1 \\
E & 6 & 191 & 8--75 & 80.9 & 2 & 2.5 & 4/4 \\
\bottomrule
\end{tabular}

%% file: tables/toplevel_change.tex
\begin{tabular}{@{}lrrrr@{}}
\toprule
Release pair & Pairs & Changed & Rate & Share of changes \\
\midrule
Major, from version zero to one & 18 & 9 & 50.0\,\% & 4.8\,\% \\
Major, later & 24 & 18 & 75.0\,\% & 9.6\,\% \\
Minor, at major version zero & 229 & 64 & 27.9\,\% & 34.2\,\% \\
Other non-major & 1,075 & 96 & 8.9\,\% & 51.3\,\% \\
All pairs & 1,346 & 187 & 13.9\,\% & 100.0\,\% \\
\bottomrule
\end{tabular}

%% file: tables/human_validation.tex
\begin{tabular}{@{}lrccl@{}}
\toprule
Announced in the notes & $n$ & Yes: A / B / M & $\kappa$: A--B / M--A / M--B & Accepted \\
\midrule
Breaking change & 48 & 11 / 13 / 14 & 0.78 / 0.84 / 0.85 & yes \\
Deprecation & 50 & 9 / 9 / 9 & 1.00 / 1.00 / 1.00 & yes \\
Module removed or moved & 48 & 8 / 4 / 14 & 0.44 / 0.54 / 0.23 & no \\
Reason given & 50 & 8 / 16 / 13 & 0.58 / 0.47 / 0.56 & no \\
\bottomrule
\end{tabular}

%% file: tables/genre_summary.tex
\begin{tabular}{@{}lrrrrrrr@{}}
\toprule
Arch. & Repos & Entries & List\,\% & Any & Maj. & GitHub & PR ref.\,\% \\
\midrule
A & 14 & 3,615 & 0.2 & 7 & 2 & 3 & 2.0 \\
B & 12 & 1,449 & 0.0 & 5 & 1 & 5 & 3.3 \\
C & 17 & 12,128 & 39.3 & 12 & 7 & 7 & 3.8 \\
D & 5 & 1,376 & 56.3 & 5 & 3 & 5 & 57.0 \\
E & 6 & 2,296 & 20.5 & 4 & 1 & 3 & 15.5 \\
\bottomrule
\end{tabular}

%% file: tables/labels_vote.tex
\begin{tabular}{@{}lrr@{}}
\toprule
Majority label & Prose (47 repos) & List (25 repos) \\
\midrule
\texttt{BREAKING}$^{\dagger}$ & 2.7 & 1.0 \\
\texttt{DEPRECATION}$^{\dagger}$ & 0.0 & 0.0 \\
\texttt{RESTRUCTURING} & 1.5 & 0.6 \\
\texttt{NEW\_CAPABILITY} & 36.5 & 28.6 \\
\texttt{BUGFIX} & 16.9 & 24.2 \\
\texttt{INTERNAL} & 19.4 & 33.3 \\
\texttt{OTHER} & 9.4 & 5.7 \\
No majority & 2.4 & 1.8 \\
\midrule
Any of the first three & 7.1 & 1.9 \\
Reason stated & 4.3 & 0.9 \\
\bottomrule
\end{tabular}

%% file: tables/rq4_funnel.tex
\begin{tabular}{@{}lr@{}}
\toprule
Step & Count \\
\midrule
Release pairs with a non-private module path removed & 195 \\
\quad with release notes & 168 \\
\quad with a structural label (2 of 3 labelers) & 67.3\,\% \\
Removed non-private module paths & 3,518 \\
\quad removal units (a removed subpackage once) & 1,211 \\
\quad named in the removing release's notes & 8.9\,\% \\
\quad named, after the hand check of the matches & 4.3\,\% \\
\quad announced as deprecated one release earlier & 2 \\
\bottomrule
\end{tabular}

%% file: tables/rq4_label_bins.tex
\begin{tabular}{@{}lrrr@{}}
\toprule
Entries in the notes & With removal & Without removal & Odds ratio \\
\midrule
1--5 & 11/40 (27.5\,\%) & 48/633 (7.6\,\%) & 4.6 \\
6--15 & 24/40 (60.0\,\%) & 65/253 (25.7\,\%) & 4.3 \\
16--40 & 31/37 (83.8\,\%) & 42/96 (43.8\,\%) & 6.6 \\
41+ & 47/51 (92.2\,\%) & 44/61 (72.1\,\%) & 4.5 \\
\bottomrule
\end{tabular}

%% file: sections/discussion.tex
\section{Discussion}\label{sec:discussion}

\paragraph{What the findings mean together} Structural change is more frequent below the top level, and the release notes rarely name it there. The notes name top-level removals about three times as often as deeper ones.

\paragraph{Relation to prior work} Our coverage rate is not directly comparable with earlier estimates, because the change unit, the documentation channels, and the coverage rule differ. Abebe et al.\ find that most release notes list 6 to 26\% of the issues a release addressed~\cite{abebe2016releasenotes}. Venturini et al.\ find 78.1\% of manifesting breaking changes in npm documented in issues, pull requests, or changelogs~\cite{venturini2023depended}. We ask whether the release's own notes name the removed module or name. This excludes documentation elsewhere and counts removals that break no caller.

\paragraph{For developers} Top-level module paths change less often than deeper ones in these histories, and disappearing import paths are candidates for review at an upgrade, with more attention at major and 0.x minor increments, which have the highest change rates and account for about half of the changes.

\paragraph{For maintainers and tool builders} The entries a user has to act on are a small share of either form, and a maintainer who publishes lists of titles can add a short prose section for them. Release-note generators can list the module paths and names a release removed, as our pipeline does.

\paragraph{For researchers} Studies that mine release notes should report the form of the notes and distinguish entry-weighted from repository-level summaries. Studies that label release notes with language models should validate code by code and at the unit the claims use.

\paragraph{Future work} A paired import study can establish how often a disappearing path fails an old expression, and a human audit of the unmatched units can replace the proxies of Section~\ref{sec:rq4}.

%% file: sections/background.tex
\section{Related Work}\label{sec:related}

\paragraph{Architectural evolution by repository mining} Le et al.\ studied architectural change across several hundred versions of 14 open-source systems~\cite{le2015empirical}, and Behnamghader et al.\ present a large-scale study of architectural evolution in open-source software systems~\cite{behnamghader2017largescale}. Later work detected architectural smells in 421 versions of eight systems~\cite{le2018empirical}, found that decay predicts issue- and change-proneness~\cite{le2021architectural}, tracked a microservice system across commits~\cite{genfer2024exploring}, and followed smells over hundreds of releases of Java~\cite{gnoyke2024evolution} and industrial C and C++ systems~\cite{sas2022evolution}, typically with Arcan~\cite{arcellifontana2017arcan}. Most of these studies analyze Java or C/C++ and several recover architectures by clustering. We measure Python import graphs directly, at the level of top-level submodules and of module files. The graphs are statically extracted rather than recovered, so their accuracy depends on package identification, parsing, and import resolution rather than on a recovery technique~\cite{garcia2013comparative}. Tool support for architectural change in Python exists as a demonstrator on single projects~\cite{rao2021ac2}.

\paragraph{Design decisions in software artifacts} Design decisions have been recovered from issues, mailing lists, commits, and pull requests~\cite{shahbazian2018recovering,soliman2021exploratory,soliman2024exploring,bhat2017automatic,zhao2024drminer,sun2026recovering}, from Stack Overflow posts and issues of AI-based systems~\cite{zhang2022aiarchdecisions}, and from commits with language models~\cite{karan2026llms}. A mapping study finds version control to be the most mined source of architectural information~\cite{jeandedieu2024mining}. We classify the decisions that maintainers announce in release documentation into a fixed taxonomy and validate the labels against two coders. We also compare the two forms in which the announcements are written.

\paragraph{Release notes} Bi et al.\ coded the release notes of 1,000 GitHub projects into information types and complemented the coding with interviews and a survey~\cite{bi2022releasenotes}. Wu et al.\ built a taxonomy of issues about release notes, in which breaking changes are the missing information that issues report most often~\cite{wu2022releasenoteissues}. Other work identifies issues, pull requests, and commits as artifacts referenced in release notes~\cite{nath2022releasenoteartifacts}, generates notes from pull requests~\cite{jiang2021deeprelease} or with language models~\cite{daneshyan2025smartnote}, and benchmarks such generators~\cite{meng2025releaseeval}. Wu et al.\ count GitHub's generator, added in October 2021, among the tools developers recommend most often, from issues about release notes rather than the notes themselves~\cite{wu2024releasenotes}. Omissions have been measured against an external list of expected changes. Abebe et al.\ find that most release notes list 6 to 26\% of the issues a release addressed~\cite{abebe2016releasenotes}, and Du and Ma find 114 of 405 verified API breaking changes in 45 Python packages missing from the changelogs~\cite{du2022aexpy}. ARENA generates release notes from the code changes of a release, deleted classes and methods included, and its evaluation found information absent from the manually written notes~\cite{moreno2017arena}. We measure that absence for module paths and names rather than generate notes. Nath et al.\ report that 47\% of the artifacts referenced in release notes lack a link to a pull request, commit, or issue~\cite{nath2025traceability}. These studies take the release note, the issue, the release, or the referenced artifact as their unit. We take the entry, aggregated per repository, and split it by form. In our searches, we found no study that treats notes produced by GitHub's built-in generator as a form of their own and compares them with prose notes. Haryono et al.\ read the changelogs of three ML libraries to collect deprecated APIs~\cite{haryono2021mlcatchup}, and Wang et al.\ find that contributors of Python libraries handle API deprecation poorly~\cite{wang2020pythondeprecated}. We label every entry of \NReposWithEntries audio-ML projects.

\paragraph{Software engineering of ML frameworks} Sculley et al.\ describe the technical debt specific to ML systems~\cite{sculley2015debt}, and Amershi et al.\ report that AI components are harder to handle as distinct modules than traditional components~\cite{amershi2019se4ml}. Studies of architecting ML-enabled systems collect challenges, practices, and design decisions of the systems that use ML~\cite{serban2022adapting,nazir2024architecting}. We study the frameworks from which such systems are built. Empirical work on deep learning frameworks and libraries covers dependency networks~\cite{han2020dldependency}, package supply chains~\cite{gao2024dlsupplychains}, code clones across releases~\cite{assi2025unraveling}, breaking changes in default arguments of data science libraries~\cite{montandon2025dabc}, assumptions in development artifacts~\cite{yang2025exploratory}, and governance~\cite{qiu2026dlgovernance}. Zhang et al.\ find API evolution patterns in six Python frameworks, among them TensorFlow and scikit-learn, that Java frameworks do not show~\cite{zhang2020pythonapi}. Breaking changes and semantic versioning have also been measured across whole package ecosystems~\cite{raemaekers2017semver,decan2021semver}, and in Maven most syntactic breaking changes arrive in non-major releases~\cite{jayasuriya2025syntactic}. A 2026 review of 97 breaking-change studies lists tooling for machine learning and data science as an open gap~\cite{chen2026breakingslr}. In our searches (arXiv, Crossref, OpenAlex, and Semantic Scholar, September and October 2026, logged in the replication package), we found no study of module evolution and release-note coverage in audio libraries. McFee et al.\ discuss open-source practice in music information retrieval without mining repositories~\cite{mcfee2019opensource}. The closest precedent in audio is a mining study of 299 open-source VST plugin repositories, which characterizes their technologies and code quality practices and does not analyze release documentation or module structure~\cite{andrei2025vst}.

\paragraph{Sustainability and dormancy} Coelho and Valente treat a project without commits for a year as a candidate for failure~\cite{coelho2017fail}, a threshold later called standard~\cite{coelho2020maintained} and used for developer abandonment~\cite{avelino2019abandonment}. Osborne et al.\ characterize company collaboration in three company-hosted AI projects~\cite{osborne2025coopetition}. Company involvement is a confound, so we code owner type and report lifecycle by owner as well as by archetype.

\paragraph{LLM-assisted annotation} Ahmed et al.\ find that model-human agreement approaches human-human agreement on some annotation tasks and not on others. They recommend that a model replace at most one human coder~\cite{ahmed2025llmannotation}. We label release-note entries of framework repositories, compare three open-weight models, and use their labels only if they pass a threshold against two human coders. Following PRIMES~\cite{demartino2025primes}, its extension~\cite{demartino2025primes2}, and recent guidelines~\cite{baltes2025llmguidelines}, we record model versions and prompts and compare the model labels with human annotations.

%% file: sections/threats.tex
\section{Threats to Validity}\label{sec:threats}

\paragraph{Construct validity} Import graphs approximate module structure. They miss dynamic imports and registry-based plugin loading, and Jin et al.\ show that the dependencies hidden by dynamic typing make Python systems look better modularized than they are~\cite{jin2023possible}, so the static representation omits some dynamic dependencies and may include conditional imports. Keeping only the most frequent package per repository does not change the result: with the releases the selector dropped and the alternative package lineages included, \PkgAllLineagesUnchangedPct\% of \PkgAllLineagesPairs pairs leave the top level unchanged against \StructPairsUnchangedPct\%, and the \PkgTransitions package switches (ESPnet, jams) are themselves undocumented top-level changes. Removed modules are detected from file paths, so a module that moves appears as a removal, and we report the share that reappears under the same name. The same holds for public names. The importability proxy of Section~\ref{sec:rq4} estimates how many removed paths were importable, and none of our measures establishes that a removal broke a caller. The name match is lexical. The proxies and the model reading of Section~\ref{sec:rq4} indicate what it may miss, and a human reading of a probability sample of the unmatched units remains to be done. The name lists come from a Python~3 parse, which rejects \SymParseErrorFiles module files at \SymParseErrorTags releases of \SymParseErrorRepos repositories, mostly Python~2 code.

Release notes record a selected subset of changes, and we claim only what maintainers document. The form rule identifies entries by form rather than by tool, so a list typed by a maintainer counts as a list. We did not check the rule by hand against a sample. Its checks are the seventeen known-answer cases and the narrower rule reported beside it. Pull requests are squashed or rebased in \NSquashRepos repositories and merged in the others, so commit counts are not fully comparable across repositories.

\paragraph{Internal validity} Archetype, owner type, and team size are confounded, and we make no causal claim about lifecycle. The lifecycle prediction was formed on the initial 17 repositories and tested on the \NAdded added ones. One author coded owner type and archetype, without knowledge of the lifecycle class, and no second coder checked them. Labels come from language models at temperature zero. The step is not deterministic across model versions, so we record model digests and release every prompt and output. The labelers agree on the code with $\kappa$ of \KappaCodeLo--\KappaCodeHi and poorly on the rationale flag (\KappaRatLo--\KappaRatHi), on which the two coders also disagree (Section~\ref{sec:validation}). The coders are authors and know the research questions, and blinding limits this influence without removing it. The coders read whole releases, so the per-entry comparison covers only the architectural codes, and the other codes rest on the model labels and on inter-model agreement. The acceptance rule accepts the models for breaking changes and deprecations and rejects them for moved modules and reasons (Table~\ref{tab:human}), so the corpus-wide shares of \texttt{RESTRUCTURING} and of stated reasons describe the models. The coded sample is \RelSheetN releases and \RelEntriesCoded entries, and we report no intervals for the $\kappa$ values. The sample caps releases at 30 entries and therefore excludes the longest releases, which carry a structural entry most often (Table~\ref{tab:rq4bins}). The 0.667 floor is the tentative threshold the guidelines state for Krippendorff's alpha, and we apply it to Cohen's $\kappa$ as a chosen heuristic.

\paragraph{External validity} The corpus covers public GitHub repositories of audio-ML packages that pass our criteria, which exclude small projects, among them many dataset, evaluation, and model repositories. Findings about archetypes describe these repositories rather than a population. The comparison cohort was chosen for a spread of owners, so it shows that a pattern also occurs outside audio and does not test whether audio differs.

\paragraph{Reliability} Several analysis choices were made after the data had been seen, and we list them so that a reader can judge their effect. Tags that are not version numbers, such as tags that ship model weights, are excluded in addition to prereleases. The form rule gained its fifth list format after we saw that four repositories end their titles in a parenthesized reference without a handle, and the narrower rule is reported alongside. The taxonomy was revised from ten codes to seven and the labeler was shown the heading after the first labeling, which is in the package. The human coding was moved from isolated entries to whole releases, half of them with a removal, after the entry sample proved to contain almost no architectural entries, and the acceptance rule gained an absolute floor. The other choices are these: archived repositories count as dormant, the classification is reported under six threshold combinations, releases are ordered by version number, structure is computed on all releases, archetype E was added after the first four, lifecycle comparisons use odds ratios with Wald intervals, \FunnelSeedsJudged seed repositories were added by hand because the topic search does not return them, and \NSeedsNotInSearch of them entered the corpus. The prerelease, heading and duplicate entries were removed after the labeling runs.

%% file: sections/conclusion.tex
\section{Conclusion}

We measure where the module structure of \StructNRepos open-source audio-ML packages changes across releases and what their release notes say about it. We find that structural change happens below the top level of a package and that the release notes rarely name it. The top level changes at a higher rate at major increments and at minor increments of packages still at version 0.x, while the other increments account for about half of the changes. Under a lexical match the notes name fewer than one removed module path or public name in ten, and longer notes and releases with module removals both carry structural entries more often. The findings concern module paths, names, and the release's own notes, not compatibility or downstream failures. Release-note generators can list the removed modules and names from the code history.

\section*{Data Availability}
The pipeline, the per-item results with provenance, the prompts and raw model outputs, the coders' answers, and the candidate ledger are in the replication package, which we intend to publish.

\section*{Acknowledgments}
An AI assistant helped draft and revise text and code. The authors reviewed all content and are responsible for it.

%% file: main.bbl
\begin{thebibliography}{10}
\providecommand{\url}[1]{#1}
\csname url@samestyle\endcsname
\providecommand{\newblock}{\relax}
\providecommand{\bibinfo}[2]{#2}
\providecommand{\BIBentrySTDinterwordspacing}{\spaceskip=0pt\relax}
\providecommand{\BIBentryALTinterwordstretchfactor}{4}
\providecommand{\BIBentryALTinterwordspacing}{\spaceskip=\fontdimen2\font plus
\BIBentryALTinterwordstretchfactor\fontdimen3\font minus
  \fontdimen4\font\relax}
\providecommand{\BIBforeignlanguage}[2]{{%
\expandafter\ifx\csname l@#1\endcsname\relax
\typeout{** WARNING: IEEEtran.bst: No hyphenation pattern has been}%
\typeout{** loaded for the language `#1'. Using the pattern for}%
\typeout{** the default language instead.}%
\else
\language=\csname l@#1\endcsname
\fi
#2}}
\providecommand{\BIBdecl}{\relax}
\BIBdecl

\bibitem{raemaekers2017semver}
S.~Raemaekers, A.~van Deursen, and J.~Visser, ``Semantic versioning and impact
  of breaking changes in the {Maven} repository,'' \emph{Journal of Systems and
  Software}, vol. 129, pp. 140--158, 2017.

\bibitem{decan2021semver}
A.~Decan and T.~Mens, ``What do package dependencies tell us about semantic
  versioning?'' \emph{IEEE Transactions on Software Engineering}, vol.~47,
  no.~6, pp. 1226--1240, 2021.

\bibitem{bi2022releasenotes}
T.~Bi, X.~Xia, D.~Lo, J.~Grundy, and T.~Zimmermann, ``An empirical study of
  release note production and usage in practice,'' \emph{IEEE Transactions on
  Software Engineering}, vol.~48, no.~6, pp. 1834--1852, 2022.

\bibitem{wu2022releasenoteissues}
J.~Wu, H.~He, W.~Xiao, K.~Gao, and M.~Zhou, ``Demystifying software release
  note issues on {GitHub},'' in \emph{Proceedings of the 30th International
  Conference on Program Comprehension (ICPC)}, 2022, pp. 602--613.

\bibitem{abebe2016releasenotes}
S.~L. Abebe, N.~Ali, and A.~E. Hassan, ``An empirical study of software release
  notes,'' \emph{Empirical Software Engineering}, vol.~21, no.~3, pp.
  1107--1142, 2016.

\bibitem{du2022aexpy}
X.~Du and J.~Ma, ``{AexPy}: Detecting {API} breaking changes in {Python}
  packages,'' in \emph{Proceedings of the 33rd IEEE International Symposium on
  Software Reliability Engineering (ISSRE)}, 2022, pp. 470--481.

\bibitem{behnamghader2017largescale}
P.~Behnamghader, D.~M. Le, J.~Garcia, D.~Link, A.~Shahbazian, and
  N.~Medvidovic, ``A large-scale study of architectural evolution in
  open-source software systems,'' \emph{Empirical Software Engineering},
  vol.~22, pp. 1146--1193, 2017.

\bibitem{genfer2024exploring}
P.~Genfer and U.~Zdun, ``Exploring architectural evolution in microservice
  systems using repository mining techniques and static code analysis,'' in
  \emph{Software Architecture. ECSA 2024}, ser. Lecture Notes in Computer
  Science, vol. 14889.\hskip 1em plus 0.5em minus 0.4em\relax Springer, 2024,
  pp. 157--173.

\bibitem{github2026releasenotesdocs}
{GitHub}, ``Automatically generated release notes,'' GitHub Docs,
  \url{https://docs.github.com/en/repositories/releasing-projects-on-github/automatically-generated-release-notes},
  2026, accessed 2026-09-27.

\bibitem{ahmed2025llmannotation}
T.~Ahmed, P.~Devanbu, C.~Treude, and M.~Pradel, ``Can {LLMs} replace manual
  annotation of software engineering artifacts?'' in \emph{Proceedings of the
  IEEE/ACM 22nd International Conference on Mining Software Repositories
  (MSR)}, 2025, pp. 526--538.

\bibitem{baltes2025llmguidelines}
S.~Baltes, F.~Angermeir, C.~Arora, M.~M. Bar{\'o}n, C.~Chen, L.~B{\"o}hme
  \emph{et~al.}, ``Guidelines for empirical studies in software engineering
  involving large language models,'' 2025.

\bibitem{jiang2021deeprelease}
H.~Jiang, J.~Zhu, L.~Yang, G.~Liang, and C.~Zuo, ``{DeepRelease}:
  Language-agnostic release notes generation from pull requests of open-source
  software,'' in \emph{Proceedings of the 28th Asia-Pacific Software
  Engineering Conference (APSEC)}, 2021, pp. 101--110.

\bibitem{venturini2023depended}
D.~Venturini, F.~R. Cogo, I.~Polato, M.~A. Gerosa, and I.~S. Wiese, ``I
  depended on you and you broke me: An empirical study of manifesting breaking
  changes in client packages,'' \emph{ACM Transactions on Software Engineering
  and Methodology}, vol.~32, no.~4, pp. 1--26, 2023.

\bibitem{le2015empirical}
D.~M. Le, P.~Behnamghader, J.~Garcia, D.~Link, A.~Shahbazian, and
  N.~Medvidovic, ``An empirical study of architectural change in open-source
  software systems,'' in \emph{Proceedings of the 12th Working Conference on
  Mining Software Repositories (MSR)}, 2015, pp. 235--245.

\bibitem{le2018empirical}
D.~M. Le, D.~Link, A.~Shahbazian, and N.~Medvidovic, ``An empirical study of
  architectural decay in open-source software,'' in \emph{IEEE International
  Conference on Software Architecture (ICSA)}, 2018.

\bibitem{le2021architectural}
D.~M. Le, S.~Karthik, M.~Schmitt~Laser, and N.~Medvidovic, ``Architectural
  decay as predictor of issue- and change-proneness,'' in \emph{IEEE 18th
  International Conference on Software Architecture (ICSA)}, 2021, pp. 92--103.

\bibitem{gnoyke2024evolution}
P.~Gnoyke, S.~Schulze, and J.~Kr{\"u}ger, ``Evolution patterns of
  software-architecture smells: An empirical study of intra- and inter-version
  smells,'' \emph{Journal of Systems and Software}, vol. 217, p. 112170, 2024.

\bibitem{sas2022evolution}
D.~Sas, P.~Avgeriou, and U.~Uyumaz, ``On the evolution and impact of
  architectural smells---an industrial case study,'' \emph{Empirical Software
  Engineering}, vol.~27, no.~4, p.~86, 2022.

\bibitem{arcellifontana2017arcan}
F.~Arcelli~Fontana, I.~Pigazzini, R.~Roveda, D.~Tamburri, M.~Zanoni, and
  E.~Di~Nitto, ``Arcan: A tool for architectural smells detection,'' in
  \emph{IEEE International Conference on Software Architecture Workshops
  (ICSAW)}, 2017, pp. 282--285.

\bibitem{garcia2013comparative}
J.~Garcia, I.~Ivkovic, and N.~Medvidovic, ``A comparative analysis of software
  architecture recovery techniques,'' in \emph{28th IEEE/ACM International
  Conference on Automated Software Engineering (ASE)}, 2013, pp. 486--496.

\bibitem{rao2021ac2}
A.~E. Rao, D.~Vagavolu, and S.~Chimalakonda, ``{AC}\textsuperscript{2}: Towards
  understanding architectural changes in {Python} projects,'' in
  \emph{Proceedings of the 29th ACM Joint European Software Engineering
  Conference and Symposium on the Foundations of Software Engineering
  (ESEC/FSE)}, 2021, pp. 1555--1559.

\bibitem{shahbazian2018recovering}
A.~Shahbazian, Y.~K. Lee, D.~Le, Y.~Brun, and N.~Medvidovic, ``Recovering
  architectural design decisions,'' in \emph{IEEE International Conference on
  Software Architecture (ICSA)}, 2018.

\bibitem{soliman2021exploratory}
M.~Soliman, M.~Galster, and P.~Avgeriou, ``An exploratory study on
  architectural knowledge in issue tracking systems,'' in \emph{Software
  Architecture. ECSA 2021}, ser. Lecture Notes in Computer Science.\hskip 1em
  plus 0.5em minus 0.4em\relax Springer, 2021, pp. 117--133.

\bibitem{soliman2024exploring}
M.~Soliman, ``Exploring architectural design decisions in mailing lists and
  their traceability to issue trackers,'' in \emph{Software Architecture. ECSA
  2024}, ser. Lecture Notes in Computer Science, vol. 14889.\hskip 1em plus
  0.5em minus 0.4em\relax Springer, 2024, pp. 307--323.

\bibitem{bhat2017automatic}
M.~Bhat, K.~Shumaiev, A.~Biesdorf, U.~Hohenstein, and F.~Matthes, ``Automatic
  extraction of design decisions from issue management systems: A machine
  learning based approach,'' in \emph{Software Architecture. ECSA 2017}, ser.
  Lecture Notes in Computer Science, vol. 10475.\hskip 1em plus 0.5em minus
  0.4em\relax Springer, 2017, pp. 138--154.

\bibitem{zhao2024drminer}
J.~Zhao, Z.~Yang, L.~Zhang, X.~Lian, D.~Yang, and X.~Tan, ``{DRMiner}:
  Extracting latent design rationale from {Jira} issue logs,'' in
  \emph{Proceedings of the 39th IEEE/ACM International Conference on Automated
  Software Engineering (ASE)}, 2024, pp. 468--480.

\bibitem{sun2026recovering}
M.~Sun, A.~Saha, N.~De~Silva, A.~Mastropaolo, and O.~Chaparro, ``Recovering
  fine-grained code change rationale from multiple software artifacts,'' 2026,
  preprint.

\bibitem{zhang2022aiarchdecisions}
B.~Zhang, T.~Liu, P.~Liang, C.~Wang, M.~Shahin, and J.~Yu, ``Architecture
  decisions in {AI}-based systems development: An empirical study,'' in
  \emph{Proceedings of the IEEE International Conference on Software Analysis,
  Evolution and Reengineering (SANER)}, 2023, pp. 616--626.

\bibitem{karan2026llms}
A.~Karan, R.~Dhar, M.~Soliman, and K.~Vaidhyanathan, ``Can {LLMs} extract
  architectural design decisions from source code commits? {A} preliminary
  exploratory study,'' in \emph{Software Architecture. ECSA 2026 Tracks and
  Workshops}, ser. Lecture Notes in Computer Science.\hskip 1em plus 0.5em
  minus 0.4em\relax Springer, 2026, pp. 208--215.

\bibitem{jeandedieu2024mining}
M.~Jean~de Dieu, P.~Liang, M.~Shahin, C.~Yang, and Z.~Li, ``Mining
  architectural information: A systematic mapping study,'' \emph{Empirical
  Software Engineering}, vol.~29, p.~79, 2024.

\bibitem{nath2022releasenoteartifacts}
S.~S. Nath and B.~Roy, ``Exploring relevant artifacts of release notes: The
  practitioners' perspective,'' in \emph{Proceedings of the IEEE International
  Conference on Software Analysis, Evolution and Reengineering (SANER)}, 2022,
  pp. 1270--1277.

\bibitem{daneshyan2025smartnote}
F.~Daneshyan, R.~He, J.~Wu, and M.~Zhou, ``{SmartNote}: An {LLM}-powered,
  personalised release note generator that just works,'' \emph{Proceedings of
  the ACM on Software Engineering}, vol.~2, no. FSE, p. FSE075, 2025.

\bibitem{meng2025releaseeval}
Q.~Meng, Z.~Ren, and J.~Visser, ``{ReleaseEval}: A benchmark for evaluating
  language models in automated release note generation,'' 2025.

\bibitem{wu2024releasenotes}
J.~Wu, H.~He, K.~Gao, W.~Xiao, J.~Li, and M.~Zhou, ``A comprehensive analysis
  of challenges and strategies for software release notes on {GitHub},''
  \emph{Empirical Software Engineering}, vol.~29, no.~5, p. 104, 2024.

\bibitem{moreno2017arena}
L.~Moreno, G.~Bavota, M.~D. Penta, R.~Oliveto, A.~Marcus, and G.~Canfora,
  ``{ARENA}: An approach for the automated generation of release notes,''
  \emph{IEEE Transactions on Software Engineering}, vol.~43, no.~2, pp.
  106--127, 2017.

\bibitem{nath2025traceability}
S.~S. Nath, B.~Roy, and M.~Jahan, ``Establishing traceability links between
  release notes and software artifacts: Practitioners' perspectives,'' in
  \emph{Proceedings of the 35th IEEE International Conference on Collaborative
  Advances in Software and Computing (CASCON)}.\hskip 1em plus 0.5em minus
  0.4em\relax IEEE, 2025, pp. 379--388.

\bibitem{haryono2021mlcatchup}
S.~A. Haryono, F.~Thung, D.~Lo, J.~Lawall, and L.~Jiang, ``Characterization and
  automatic updates of deprecated machine-learning {API} usages,'' in
  \emph{Proceedings of the IEEE International Conference on Software
  Maintenance and Evolution (ICSME)}, 2021, pp. 137--147.

\bibitem{wang2020pythondeprecated}
J.~Wang, L.~Li, K.~Liu, and H.~Cai, ``Exploring how deprecated {Python} library
  {APIs} are (not) handled,'' in \emph{Proceedings of the 28th ACM Joint
  Meeting on European Software Engineering Conference and Symposium on the
  Foundations of Software Engineering (ESEC/FSE)}, 2020, pp. 233--244.

\bibitem{sculley2015debt}
D.~Sculley, G.~Holt, D.~Golovin, E.~Davydov, T.~Phillips, D.~Ebner
  \emph{et~al.}, ``Hidden technical debt in machine learning systems,'' in
  \emph{Advances in Neural Information Processing Systems 28 (NIPS)}, 2015.

\bibitem{amershi2019se4ml}
S.~Amershi, A.~Begel, C.~Bird, R.~DeLine, H.~Gall, E.~Kamar \emph{et~al.},
  ``Software engineering for machine learning: A case study,'' in
  \emph{Proceedings of the IEEE/ACM 41st International Conference on Software
  Engineering: Software Engineering in Practice (ICSE-SEIP)}, 2019, pp.
  291--300.

\bibitem{serban2022adapting}
A.~Serban and J.~Visser, ``Adapting software architectures to machine learning
  challenges,'' in \emph{Proceedings of the IEEE International Conference on
  Software Analysis, Evolution and Reengineering (SANER)}, 2022, pp. 152--163.

\bibitem{nazir2024architecting}
R.~Nazir, A.~Bucaioni, and P.~Pelliccione, ``Architecting {ML}-enabled systems:
  Challenges, best practices, and design decisions,'' \emph{Journal of Systems
  and Software}, vol. 207, p. 111860, 2024.

\bibitem{han2020dldependency}
J.~Han, S.~Deng, D.~Lo, C.~Zhi, J.~Yin, and X.~Xia, ``An empirical study of the
  dependency networks of deep learning libraries,'' in \emph{Proceedings of the
  IEEE International Conference on Software Maintenance and Evolution (ICSME)},
  2020, pp. 868--878.

\bibitem{gao2024dlsupplychains}
K.~Gao, R.~He, B.~Xie, and M.~Zhou, ``Characterizing deep learning package
  supply chains in {PyPI}: Domains, clusters, and disengagement,'' \emph{ACM
  Transactions on Software Engineering and Methodology}, vol.~33, no.~4, pp.
  1--27, 2024.

\bibitem{assi2025unraveling}
M.~Assi, S.~Hassan, and Y.~Zou, ``Unraveling code clone dynamics in deep
  learning frameworks,'' \emph{ACM Transactions on Software Engineering and
  Methodology}, vol.~34, no.~8, pp. 1--39, 2025.

\bibitem{montandon2025dabc}
J.~E. Montandon, L.~L. Silva, C.~Politowski, D.~Prates,
  A.~de~Brito~Bonif{\'a}cio, and G.~E. Boussaidi, ``Unboxing default argument
  breaking changes in 1 + 2 data science libraries,'' \emph{Journal of Systems
  and Software}, vol. 229, 2025.

\bibitem{yang2025exploratory}
C.~Yang, P.~Liang, and Z.~Ma, ``An exploratory study on automatic
  identification of assumptions in the development of deep learning
  frameworks,'' \emph{Science of Computer Programming}, vol. 240, p. 103218,
  2025.

\bibitem{qiu2026dlgovernance}
R.~Qiu, W.~Yang, and M.~Zhou, ``From collaboration to regulation:
  Characterizing governance practice in three deep learning open source
  communities,'' 2026.

\bibitem{zhang2020pythonapi}
Z.~Zhang, H.~Zhu, M.~Wen, Y.~Tao, Y.~Liu, and Y.~Xiong, ``How do {Python}
  framework {APIs} evolve? {An} exploratory study,'' in \emph{Proceedings of
  the 27th IEEE International Conference on Software Analysis, Evolution and
  Reengineering (SANER)}, 2020, pp. 81--92.

\bibitem{jayasuriya2025syntactic}
D.~Jayasuriya, S.~Ou, S.~Hegde, V.~Terragni, J.~Dietrich, and K.~Blincoe, ``An
  extended study of syntactic breaking changes in the wild,'' \emph{Empirical
  Software Engineering}, vol.~30, no.~2, p.~42, 2025.

\bibitem{chen2026breakingslr}
J.~Chen, T.~Bi, Y.~Wang, and P.~Thongtanunam, ``Breaking changes in software
  ecosystems: A systematic literature review,'' arXiv:2605.24397, 2026,
  preprint.

\bibitem{mcfee2019opensource}
B.~McFee, J.~W. Kim, M.~Cartwright, J.~Salamon, R.~Bittner, and J.~P. Bello,
  ``Open-source practices for music signal processing research: Recommendations
  for transparent, sustainable, and reproducible audio research,'' \emph{IEEE
  Signal Processing Magazine}, vol.~36, no.~1, pp. 128--137, 2019.

\bibitem{andrei2025vst}
B.~Andrei, M.~V. Merino, and I.~Malavolta, ``The ecosystem of open-source music
  production software -- a mining study on the development practices of {VST}
  plugins on {GitHub},'' in \emph{Proceedings of the IEEE/ACM 22nd
  International Conference on Mining Software Repositories (MSR)}, 2025, pp.
  513--525.

\bibitem{coelho2017fail}
J.~Coelho and M.~T. Valente, ``Why modern open source projects fail,'' in
  \emph{Proceedings of the 11th Joint Meeting on Foundations of Software
  Engineering (ESEC/FSE)}, 2017, pp. 186--196.

\bibitem{coelho2020maintained}
J.~Coelho, M.~T. Valente, L.~Milen, and L.~L. Silva, ``Is this {GitHub} project
  maintained? measuring the level of maintenance activity of open-source
  projects,'' \emph{Information and Software Technology}, vol. 122, p. 106274,
  2020.

\bibitem{avelino2019abandonment}
G.~Avelino, E.~Constantinou, M.~T. Valente, and A.~Serebrenik, ``On the
  abandonment and survival of open source projects: An empirical
  investigation,'' in \emph{Proceedings of the ACM/IEEE International Symposium
  on Empirical Software Engineering and Measurement (ESEM)}, 2019, pp. 1--12.

\bibitem{osborne2025coopetition}
C.~Osborne, F.~Daneshyan, R.~He, H.~Ye, Y.~Zhang, and M.~Zhou, ``Characterising
  open source co-opetition in company-hosted open source software projects: The
  cases of {PyTorch}, {TensorFlow}, and {Transformers},'' \emph{Proceedings of
  the ACM on Human-Computer Interaction}, vol.~9, no.~2, pp. 1--30, 2025.

\bibitem{demartino2025primes}
V.~D. Martino, J.~Casta{\~n}o, F.~Palomba, X.~Franch, and
  S.~Mart{\'i}nez-Fern{\'a}ndez, ``A framework for using {LLMs} for repository
  mining studies in empirical software engineering,'' in \emph{Proceedings of
  the IEEE/ACM International Workshop on Methodological Issues with Empirical
  Studies in Software Engineering (WSESE)}, 2025, pp. 6--11.

\bibitem{demartino2025primes2}
------, ``A methodological framework for {LLM}-based mining of software
  repositories,'' 2025.

\bibitem{jin2023possible}
W.~Jin, D.~Zhong, Y.~Cai, R.~Kazman, and T.~Liu, ``Evaluating the impact of
  possible dependencies on architecture-level maintainability,'' \emph{IEEE
  Transactions on Software Engineering}, vol.~49, no.~3, pp. 1064--1085, 2023.

\end{thebibliography}
